\documentclass[useAMS,usenatbib]{mn2e}
\usepackage{epsf}
\newcommand{\simgt}{\lower.5ex\hbox{$\; \buildrel > \over \sim \;$}}
\newcommand{\simlt}{\lower.5ex\hbox{$\; \buildrel < \over \sim \;$}}
\newcommand{\rth}{R_{\rm th}}
\newcommand{\sigmam}{\sigma_{\rm\scriptscriptstyle M}}

\newcommand{\rhosm}{\rho_{\rm\scriptscriptstyle smooth}}
\newcommand{\bfr}{\mbox{\boldmath$r$}}
\newcommand{\rhopower}{\rho_{\rm\scriptscriptstyle power}}
\newcommand{\rhonfw}{\rho_{\rm\scriptscriptstyle NFW}}

\newcommand{\delzero}{\delta_0}
\newcommand{\Dvir}{\Delta_{\rm vir}}
\newcommand{\yr}{y_{\rm\scriptscriptstyle R }}
\begin{document}
\title{Non-Gaussian Tails of Cosmological Density Distribution 
  Function from Dark Halo Approach}
%
%
%
\date{\today}
\pubyear{2002} \volume{000} \pagerange{1} \onecolumn
%
%
%
\author[A. Taruya, T. Hamana and  I. Kayo]
{Atsushi Taruya$^1$, Takashi Hamana$^2$ and Issha Kayo$^3$\\
$^1$ Research Center for the Early Universe(RESCEU), School of Science, 
University of Tokyo, Tokyo 113-0033, Japan\\
$^2$ National Astronomical Observatory, Mitaka, Tokyo 181-8588, Japan\\ 
$^3$ Department of Physics, University of Tokyo, Tokyo 113-0033, Japan }
\maketitle \label{firstpage}
%
%
%
%
%
%
%
%
%
%
\begin{abstract}
We present a simple model based on the dark halo approach which provides
a useful way to understand key points determining the shape of the 
non-Gaussian tails of the dark matter one-point probability distribution 
function(PDF). In particular, using the scale-free models with power-law 
profile of dark halos, we derive a simple analytic expression for the 
one-point PDF. 
It is found that the shape of the PDF changes at the characteristic
value of $\delta_*$ which is defined by the smoothed density of a halo 
with the characteristic mass $M_*$ at the epoch. 
In cold dark matter models with top-hat smoothing filters, 
the characteristic smoothed density at present time typically 
takes the value $\delta_*\gg 1$ for a small smoothing 
scale $\rth\sim 1$Mpc$/h$ and conversely $\delta_*\ll 1$ for a 
large smoothing scale $\rth > 10$Mpc$/h$. On the range 
$\delta/\delta_*<1$, the shape of the PDF is almost solely 
determined by the outer slope of halos and scales as a power-law.
The resultant non-Gaussian tails of PDF then resemble the 
log-normal PDFs in that range and show a good agreement with N-body 
simulations, which can be ascribed to the universality 
of the outer slope of the halo profile. 
In contrast, tails of one-point PDF in the range $\delta/\delta_*>1$ 
basically follow the steep exponential tails of the halo mass function, 
which exhibit a strong sensitivity to both the outer slope of the halo 
profile and the initial power spectrum. 
Based on these results, the discussion  
on the PDF of galaxy distribution and the application to the weak 
lensing statistics are also presented.  
\end{abstract}
%
%
%
%
%
%
%
%
%
%
%
%
%
%
\section{Introduction}
\label{sec: intdo}
%
%
%
%
The probability distribution function (PDF) of the cosmic matter 
density is one of most fundamental quantities characterizing 
statistical properties of large-scale structures.
In most standard scenarios of the structure formation, 
the large-scale structures are thought to form by gravitational 
instabilities from small Gaussian density fluctuations.
A deviation in the PDF from the Gaussian distribution is thus
most likely caused by the nonlinear gravitational evolution
and is a key to understand the structure formation history.
The higher order moments of the density field such like the 
skewness and kurtosis have been used to study the 
non-Gaussian nature in cosmic density field (e.g., Peebles 1980; 
Juszkiewicz, Bouchet \& Colombi 1993; Bernardeau 1994a). 
The one-point PDF contains more statistical information
than them.

The galaxy PDFs have been measured from various catalogs (e.g., Hamilton 1985; Bouchet et al. 1993; Kofman et al. 1994) 
and shows significant non-Gaussianity. 
Indeed it was pointed out that the galaxy PDF is well fitted by 
the log-normal distribution (e.g., Coles \& Jones 1991; Kofman et al. 1994).
The dark matter PDFs measured from $N$-body simulations are also 
fitted by the log-normal distribution very well (e.g., Bernardeau \& Kofman 
1995; Taylor \& Watts 2000). 
Although the galaxy distribution cannot be regarded as a direct tracer 
of the dark matter distribution because of unknown biasing relation 
between them, the similarity in their PDF suggests that there exists a 
nature common to their distributions.

The origin of the phenomenological log-normal PDF has not yet been clarified. 
Bernardeau (1994b) and Benardeau \& Kofman (1995) argued that the 
perturbative construction of one-point PDF shows a strong dependence on 
the primordial power spectrum and the PDF approaches the log-normal form 
only when the initial spectrum is proportional to $P(k)\propto k^{n}$ 
with $n=-1$. On the other hand, using a series of high-resolution 
N-body simulations, Kayo, Taruya \& Suto (2001) found that the one-point 
density PDFs with Gaussian initial 
conditions are approximately described by the log-normal distribution in 
the strongly non-linear regime. Remarkably, the accuracy 
of log-normal model turns out to become almost 
insensitive to the underlying power spectrum of density fluctuation, 
in contrast to the prediction from perturbation theory. 
While the results of Kayo et al. (2001) show a clear dependence on the 
primordial initial spectrum in a weakly non-linear regime, consistent 
with the perturbation theory, there remains crucial to clarify the reason 
why the PDFs in the simulations tend to become insensitive to the initial 
power spectrum and approach to the log-normal form in a strongly 
non-linear regime.

The primary purpose of this paper is to discuss this issue using 
a simple analytic model. That is, employing the so-called {\it dark 
halo approach} as advocated recently, we construct a model for 
the non-Gaussian tails of the one-point PDF, which includes 
the essential ingredient for gravitational clustering and is, at least, 
consistent with N-body simulations. 
The basic assumption of our model is that the high density regions after 
smoothed over with a filter function are mainly attributed to a single 
massive halo. Clearly, this assumption is violated at the low 
density regions and we thus restrict our analysis to the highly 
non-Gaussian tails of mass density PDF. 
Then, the dark halo approach allows us to derive a simple 
analytic expression for the one-point PDF in cases with the 
scale-free initial spectra and power-law profiles of dark halo, 
from which we explain the qualitative behavior of tails of PDF 
and its dependences on the halo density profile and on 
the initial power spectrum. 
Further, adopting the realistic halo profile and the mass 
function, the results of our analytic model are quantitatively 
compared with N-body simulations with various initial power spectra.

In section \ref{sec: halo_approach}, the basic ingredients for 
the dark halo approach and an analytic model of tails of one-point PDF 
are described. Section \ref{sec: results} presents the detailed 
investigation of the dependence on the halo profile and the sensitivity to the 
initial power spectra 
on the non-Gaussian tails of one-point PDF. 
Assuming the scale-free initial spectrum and the 
power-law profile of dark halos, we analytically derive an approximate 
expression 
for non-Gaussian tails of PDF. Using this result, we attempt to clarify the 
sensitivity of the one-point PDF to the initial spectra and the halo profiles. 
In section \ref{sec: comparison}, the analytic model of 
one-point PDF is quantitatively compared with those obtained from cosmological 
N-body simulations. Finally, section \ref{sec: conclusion} is devoted to 
the conclusions and the discussion. 
%
%
%
%
%
%
%
%
%
%
%
\section{Mass density PDF from the dark halo approach}
\label{sec: halo_approach}
%
%
%
%
\subsection{Basic ingredients}
%
%
%
%
%
%
The basic idea of the dark halo approach which has become very popular 
recently(e.g., Seljak 2000; Ma \& Fry 2000a; Scoccimarro et al. 2001)
is that (i) all the dark matter in the universe belong to one of the 
virialized clumps (dark halos), 
and (ii) the matter distribution in the universe is approximated by 
a sum of the matter distribution within each halo.
To construct an analytic model of mass density PDF from the dark 
halo approach, we need to specify the halo mass function and the 
density profiles of halos. In the following, we describe these two 
ingredients.

For the halo mass function, the approximate prescription for 
counting the number of dark halos has been first presented by 
Press \& Schechter (1974) and recently an accurate analytic fitting 
model has been 
proposed by Sheth \& Tormen (1998). Both models are simply expressed 
in a unified manner as follows: 
\begin{equation}
n(M)dM=A\left\{1+\left(\frac{\sigmam}
{\sqrt{a}\,\delta_{\rm crit}}\right)^{2p}\right\}\sqrt{\frac{a}{2\pi}}\,\,
\frac{\overline\rho}{M}\frac{\delta_{\rm crit}}{\sigmam^3}
\exp\left\{-\frac{a}{2}\left(\frac{\delta_{\rm crit}}{\sigmam}\right)^2\right\}\,
\left|\frac{d\sigmam^2}{dM}\right|dM,
\label{eq: mass_function}
\end{equation}
where $\overline\rho$ is the mean mass density and $\delta_{\rm crit}$ 
is the critical over-density of the spherical collapse model(see Nakamura 
\& Suto 1997 for useful fitting formula). The quantity 
$\sigmam$ denotes the linear rms fluctuations of the matter density in a 
sphere of radius $R_M=(3M/4\pi\overline\rho)^{1/3}$. The numerical 
coefficients $a$ and $p$ are set to $a=1$, $p=0$ for the Press--Schechter 
mass function and  
$a=0.707$, $p=0.3$ for the mass function by Sheth \& Tormen. The 
normalization factor $A$ is determined by $\int n(M) M dM=\overline\rho$. 
Throughout the paper, we extensively use the mass function by Sheth \& 
Tormen (1998).

The density profile of dark halos is defined by a spherical average 
for each halo and is given by a function of radius and halo mass.  
The numerical simulation by Navarro et al. (1996,1997; hereafter NFW) 
revealed that the outer slope of profile becomes $-3$, while the inner 
part asymptotically approaches $-1$. In this paper, we adopt their  
fitting form: 
\begin{equation}
\frac{\rhonfw(r;M)}{\overline\rho}=\frac{\delta_c}{(r/r_s)(1+r/r_s)^2},
        \label{eq: NFW_profile}
\end{equation}
where $\delta_c$ and $r_s$ respectively denote the characteristic over-density 
and the scale radius. These parameters are not independent and are 
characterized in terms of concentration parameter $c\equiv r_{\rm vir}/r_s$, 
with $r_{\rm vir}$ being virial radius of the halo. 
Using this quantity, the characteristic over-density 
$\delta_c$ is given by $\delta_c=(\Dvir/3) c^3/[\ln(1+c)-c/(1+c)]$, 
where $\Dvir$ denotes the virial over-density of spherical collapse 
model (Nakamura \& Suto 1997). 
For the relation between the concentration parameter and the halo mass, 
we adopt the following fitting form according to the simulation by 
Bullock et al. (2001), 
\begin{equation}
c(M)= c_*\,\,\left(\frac{M}{M_*}\right)^{-\beta}, 
\label{eq: c(M)}
\end{equation}
where $M_*$ corresponds to the characteristic mass scale satisfying 
$\sigmam(M_*)=\delta_{\rm crit}$. Below in comparing the PDF with 
the N-body simulation (Sec.\ref{sec: comparison}), 
the parameters $c_*$ and $\beta$ are chosen 
as $(c_*,\beta)=(10,0.15)$ for cold dark matter(CDM) model and  $(10,(n+3)/6)$ for 
scale-free model with spectral index $n$. 
The systematic influences of the uncertainty of $c(M)$ on the evaluation of PDF 
are discussed in detail in Appendix B.

Indeed, recent high-resolution N-body study 
reveals that the inner slope of halo profile might not converge to $-1$, 
but it rather approaches $-1.5$ (Moore et al. 1999; 
Fukushige \& Makino 2001a,b; see also Jing \& Suto 2000). 
Hence, it becomes important to clarify the profile dependence on the 
non-Gaussian tails of PDF. In this paper, in addition to the NFW profile, 
we further treat the halo models characterized by a single power-law 
profile: 
\begin{equation}
\frac{\rhopower(r;M)}{\overline\rho}=\delta_c\,
        \left(\frac{r}{r_s}\right)^{-\alpha},~~~~~(0<\alpha<3)
        \label{eq: power-law_profile}
\end{equation}
which enables us to derive the analytic expressions of 
PDFs (see Sec.\ref{sec: results}). Here, the characteristic density and scale 
radius are determined by the relation between virial mass and virial radius:  
\begin{equation}
\delta_c\,r_s^{\alpha} = \frac{3-\alpha}{3}\,\,\Dvir\,r_{\rm vir}^{\alpha}
        =\frac{3-\alpha}{3}\,\Dvir\,
\left(\frac{3M}{4\pi\,\overline\rho\Dvir}\right)^{\alpha/3}.
\label{eq: delc_rs_power}
\end{equation}

%
%
%
%
%
\subsection{Mass density PDFs}
\label{subsec: mass_density_PDFs}
%
%
%
%
Provided the basic ingredients, we now describe a simple analytic 
model of mass density PDF. 

Let us first describe the cumulative distribution function 
$P(>\delta;R)$, which gives a probability that a given density field 
smoothed over the radius $R$ is larger than $\delta$. 
Defining the smoothed halo profile:  
\begin{equation}
\rhosm(r;R,M)=\int d^3\bfr'\,\,
        W(|\bfr-\bfr'|;R)\,\,\rho(r';M), 
\label{eq: rho_smooth}
\end{equation}
with the function $W(x;R)$ being the spherically symmetric 
smoothing kernel.
Since we only consider simple spherical symmetric density 
distributions of halos, one can calculate the characteristic 
radius $r(\delta;R,M)$ by solving the following equation: 
\begin{equation}
\frac{\rhosm[r(\delta;R,M);R,M]}{\overline\rho}
=\delta + 1\simeq \delta. 
\label{eq: def_of_r_delta}
\end{equation}
Now we introduce the volume factor $V(>\delta; R,M)$
which represents, for one halo with mass $M$, the comoving 
volume in which the smoothed mass density field is always 
greater than $\delta$.
Since the smoothed halo profile is a monotonic decreasing function 
of the radius, the volume factor is simply given by, 
\begin{equation}
V(>\delta; R,M)=\frac{4\pi}{3}\left\{r(\delta;R,M)\right\}^3. 
\label{eq: volume_factor}
\end{equation}

The volume factor (\ref{eq: volume_factor}) explicitly depends on the 
halo mass and thereby the volume fraction of the entire region greater 
than $\delta$ is evaluated from the integration of 
equation (\ref{eq: volume_factor}) 
over the different halo mass weighting the comoving number density of  
dark halos. Indeed, this volume fraction is equivalent to 
the cumulative distribution function $P(>\delta;R)$. We thus obtain:  
\begin{equation}
P(>\delta; R) = \int_0^{\infty}\,dM\,\,V(>\delta;R,M)\,n(M). 
\label{eq: cdf}
\end{equation}
Once provided the cumulative distribution, it is easy to calculate 
the one-point PDF from the definition 
$P(>\delta;R) = \int_{\delta}^{+\infty}d\delta\,P(\delta;R)$: 
\begin{equation}
P(\delta;R) = \left|\,\frac{d}{d\delta}\,P(>\delta; R) \right|.
\label{eq: pdf}
\end{equation}

Equation (\ref{eq: pdf}) with equation (\ref{eq: cdf}) is the 
heart of our subsequent analysis. 
It provides a simple analytic estimate for the degree of non-Gaussian 
tails, which is, in principle applicable to  both weakly non-linear 
and strongly non-linear regimes. 
Note that the expression (\ref{eq: cdf}) heavily relies on 
the assumption that any of the dark matter particles reside in each 
dark halo. 
This would be true for high density parts of the mass density PDF, but might 
not be the case for the low density regions, $-1<\delta\simlt1$.  
We shall therefore focus on the high density tails, $\delta\gg1$. 
As shown in the following sections, this approach is very useful to 
understand the essential physical mechanisms that determine the behavior 
of the high density tail of the PDF.

While the above model prescription is general and does not restrict the 
choice of the smoothing filter, hereafter, we specifically treat 
the density field smoothed over the top-hat filter function: 
\begin{equation}
W_{\rm th}(\bfr;\rth)=\frac{3}{4\pi \rth^3}\,\,\Theta(\rth-|\bfr|), 
\label{eq: top-hat_filter}
\end{equation}
where the quantity $\Theta(x)$ denotes Heaviside step function. 
%
%
%
%
%
%
%
%
%
\section{Dependences of power spectrum index and halo profile}
\label{sec: results}
%
%
%
Before addressing the comparison with N-body simulations, 
it is instructive to consider how the halo density profile and initial 
power spectrum alter the non-Gaussian tails of PDF in a simple 
analytical manner. In this section, simply 
assuming the power-law model of halo 
density profile (\ref{eq: power-law_profile}) and scale-free initial 
spectra $P(k)\propto k^n$, 
we here derive an approximate expression for the one-point PDF 
$P(\delta;\rth)$.

To compute the PDFs, one must first evaluate the volume 
factor (\ref{eq: volume_factor}), which is given in terms of 
the characteristic radius $r(\delta;\rth,M)$. 
In Appendix A, the exact expressions for top-hat smoothed density 
profile is presented in the case of power-law profiles 
(eqs.[\ref{eq: smooth_halo_1}][\ref{eq: smooth_halo_2}]). 
The resultant smoothed profiles 
become constant at an inner region $r\ll\rth$ and show 
the power-law behavior at $r\gg\rth$ (see left panel of 
Fig.\ref{fig: rho_smooth_power}).  
Based on these exact results, in Appendix A, we further obtain the 
approximate expression for the 
smoothed density profiles, which enable us to calculate the 
quantity $r(\delta;\rth,M)$ analytically (see eq.[\ref{appen_A: approx}]):  
\begin{equation}
\displaystyle
\frac{\rhosm(r;\rth,M)}{\overline\rho}\simeq
\left\{
\begin{array}{lcr}
\delzero(\rth;M) & ; & 
        r<\left(\frac{3-\alpha}{3}\right)^{1/\alpha}\,\rth
\\
\\
\left(\frac{3-\alpha}{3}\right)\, 
\left(\frac{r}{\rth}\right)^{-\alpha} \delzero(\rth;M) & ; & 
        r\geq\left(\frac{3-\alpha}{3}\right)^{1/\alpha}\,\rth
\end{array}
\right. , 
\label{eq: rho_power_approx}
\end{equation}
where the function $\delzero(\rth,M)$ represents 
the maximum value of the smoothed density profile, 
$\rhosm(0;\rth,M)/\overline\rho$, and is given by a function of halo mass: 
\begin{equation}
\delzero(\rth,M)\,
\equiv \Dvir\,\left(\frac{\rth}{r_{\rm vir}}\right)^{-\alpha}
= \Dvir\,\left(\frac{M_R}{M}\right)^{-\alpha/3}, 
\label{eq: def_of_delpow}
\end{equation}
The quantity $M_R$ means the effective halo mass inside the smoothing 
radius $\rth$, $M_R\equiv(4\pi/3)\overline\rho\Dvir\,\rth^3$. 
For a given smoothing length $\rth$, equation (\ref{eq: def_of_delpow}) 
relates the mass of a halo with its characteristic smoothed density. 
Adopting the above approximation (\ref{eq: rho_power_approx}), 
one can easily solve (\ref{eq: def_of_r_delta}), which yields 
\begin{equation}
\displaystyle
r(\delta;\rth,M)=\left\{
\begin{array}{lcr}
0 & ; & \delta > \delzero(\rth;M)
\\
\\
\left\{\frac{3-\alpha}{3}\,
\frac{\delzero(\rth,M)}{\delta}\right\}^{1/\alpha}\,\rth & ; & 
        \delta \leq \delzero(\rth;M)
\end{array}
\right. .
\label{eq: r_delta_pow}
\end{equation}
Then, with a help of (\ref{eq: def_of_delpow}), 
substitution of equation (\ref{eq: r_delta_pow}) into 
(\ref{eq: volume_factor}) leads to 
\begin{equation}
V(>\delta;\rth,M)=\left\{
\begin{array}{lcr}
0 & ; & \delta > \delzero(\rth;M)
\\
\\
\left(\frac{3-\alpha}{3}\,\,\frac{\Dvir}{\delta}\right)^{3/\alpha} 
        \frac{M}{\overline\rho\,\Dvir} & ; & 
        \delta \leq \delzero(\rth;M)
\end{array}
\right. . 
\label{eq: volume_factor_pow}
\end{equation}
Clearly, the volume factor decreases as $\delta^{-3/\alpha}$ 
and is cut off at the maximum density $\delzero(\rth;M)$.
This indicates that for a certain value of $\delta$, the dominant 
contribution to the integral in cumulative distribution $P(>\delta;\rth)$ 
is mainly attributed to the halos whose maximum density is larger than that.  

Now, turn to focus on the evaluation of the integral in 
(\ref{eq: cdf}). For the cosmological models with initial 
scale-free spectra, the rms fluctuation $\sigmam$ in 
(\ref{eq: mass_function}) can be expressed as 
\begin{equation}
  \label{eq: rms_fluc}
  \sigmam\,\,=\,\,\delta_{\rm crit}\,\left(\frac{M}{M_*}\right)^{-(n+3)/6}. 
\end{equation}
Substituting the mass function 
(\ref{eq: mass_function}) and the volume factor 
(\ref{eq: volume_factor_pow}) into (\ref{eq: cdf}), 
the cumulative distribution function is analytically evaluated as follows:
\begin{eqnarray} 
P(>\delta;\,\rth) 
&=& \left(\frac{3-\alpha}{3}\frac{\Dvir}{\delta}\right)^{3/\alpha}\,
\frac{1}{\overline\rho\,\Dvir}\, 
\int_{(\delta/\Dvir)^{3/\alpha}M_R}^{+\infty}\,\,
n(M)\,M\,dM \nonumber\\
&=& \left(\frac{3-\alpha}{3}\frac{\Dvir}{\delta}\right)^{3/\alpha}\,
G(\delta;\alpha,n),
\label{eq: cdf_power}
\end{eqnarray} 
where 
\begin{eqnarray}
G(\delta;\alpha,n)
&\equiv& 
\frac{1}{\overline\rho\,\Dvir}\, 
\int_{(\delta/\Dvir)^{3/\alpha}M_R}^{+\infty}\,\,
n(M)\,M\,dM \nonumber\\
&=&
\frac{A}{\Dvir\sqrt{\pi}}\,\,
\int_{\yr(\delta)/2}^{+\infty}\left\{1+\frac{1}{(2x)^p}\right\}
x^{1/2}\,e^{-x}\,\frac{dx}{x} \nonumber\\
& = &
\frac{A}{\Dvir\sqrt{\pi}}\,\left\{
\Gamma\left(\frac{1}{2},\frac{\yr(\delta)}{2}\right)
+ 2^{-p}\,\,\Gamma\left(\frac{1}{2}-p,\frac{\yr(\delta)}{2}\right)\right\}
\label{eq: G}
\end{eqnarray}
with the function $\Gamma(a,x)$ being the incomplete Gamma function, 
$\Gamma(a,x)\equiv \int_{x}^{\infty}dt\,t^{a-1} e^{-t}$. 
Here, the function $\yr(\delta)$ is defined by 
\begin{equation}
\yr(\delta)\equiv a\,\left(\frac{M_R}{M_*}\right)^{(n+3)/3}
\left(\frac{\delta}{\Dvir}\right)^{(n+3)/\alpha}
= a\,\left(\frac{\delta}{\delta_*}\right)^{(n+3)/\alpha},
\label{eq: y-function}
\end{equation}
where $\delta_*\equiv\delzero(\rth,M_*)$ and denotes 
the characteristic smoothed density of the halo with mass $M_*$, 
which play a key role to reveal the sensitivity of the non-Gaussian tails. 

Equation (\ref{eq: cdf_power}) indicates that a shape of the 
PDF is determined by the combination of two contributions:
One is the halo profile. The scaling $\delta^{-3/\alpha}$ 
in equation (\ref{eq: cdf_power}) reflects the 
slope of the halo profile. The flatter the halo profile is, 
the more rapidly the volume factor varies, thus the steeper 
the PDF becomes.
Another contribution is the shape of the mass function, since the factor 
$M/\overline\rho\Dvir$ gives the characteristic volume of 
the halo with the mass $M$, and the lower limit of the
integration in equation (\ref{eq: G}) is, roughly speaking, 
determined by the halo mass whose characteristic smoothed 
density (defined by eq.[\ref{eq: def_of_delpow}]) just coincides with 
$\delta$. Therefore, the function $G(\delta;\alpha,n)$ gives the 
volume fraction of halos (i.e., a sum of characteristic 
volume of halos) whose mass is larger than such lower limit,
and directly reflects the shape of the halo mass function.
Note that the dependence of the spectral index $n$ on the PDF
enters only through the function $G(\delta;\alpha,n)$.

\begin{figure}
  \begin{center}
\begin{minipage}{0.49\textwidth}
    \epsfxsize=8.0cm
    \epsfbox{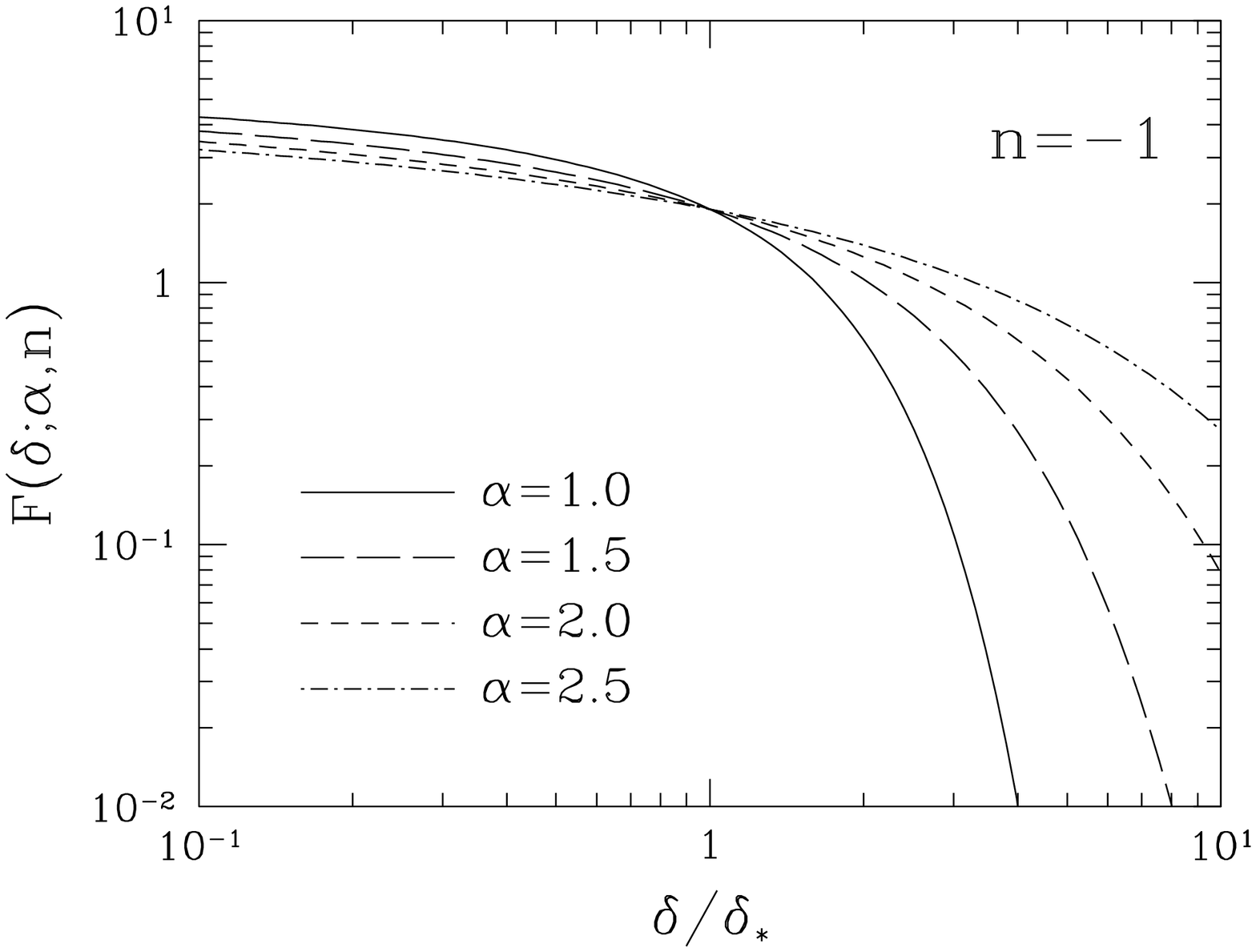}
\end{minipage}
\begin{minipage}{0.49\textwidth}
    \epsfxsize=8.0cm
    \epsfbox{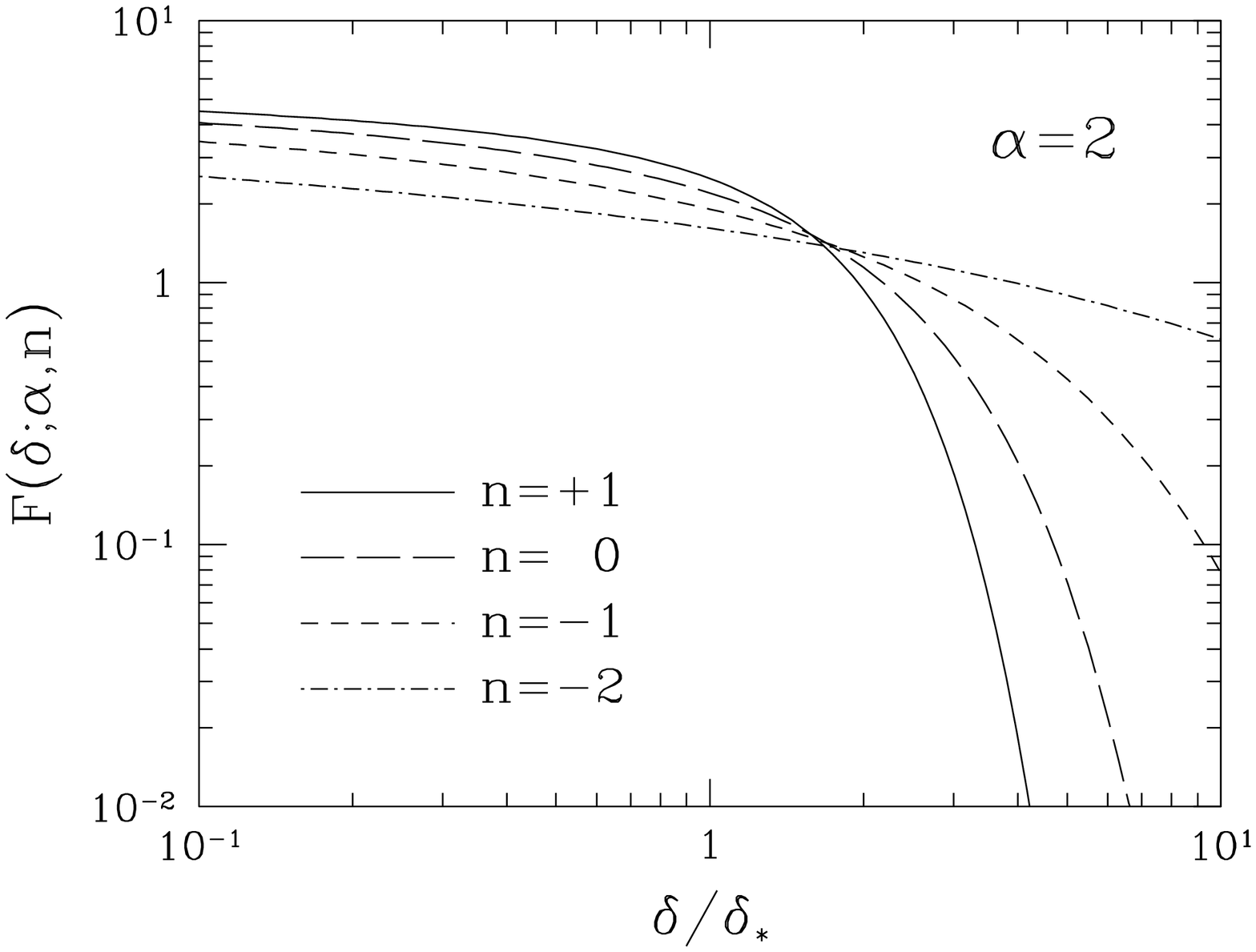}
\end{minipage}
\end{center}
    \caption{$F(\delta,\alpha,n)$ defined by eq.(\ref{eq: F}) 
as a function of $\delta/\delta_*$. 
Dependence of the halo density profiles ({\it left}) 
and initial spectra({\it right}). \label{fig: ff}}
\end{figure}

Now from (\ref{eq: cdf_power}), the one-point PDF 
$P(\delta;\rth)$ is analytically expressed as follows: 
\begin{eqnarray}
P(\delta;\rth)&=&\frac{3}{\alpha}\,
\left(\frac{3-\alpha}{3}\right)^{3/\alpha}
\frac{1}{\Dvir}\,
\left(\frac{\delta}{\Dvir}\right)^{-(3/\alpha+1)}\,
F(\delta;\alpha,n),
\label{eq: pdf_power}
\end{eqnarray}
with
\begin{eqnarray}
F(\delta;\alpha,n)
&=&
G(\delta;\alpha,n)+\frac{\alpha}{3}\Dvir\,\,
\left|\frac{d}{d\delta}G(\delta;\alpha,n)\right|
\nonumber\\
&=&
\frac{A}{\Dvir\sqrt{\pi}}\,
\left[\Gamma\left(\frac{1}{2},\frac{\yr(\delta)}{2}\right) + 2^{-p}\,
\Gamma\left(\frac{1}{2}-p,\frac{\yr(\delta)}{2}\right)+\frac{n+3}{3\sqrt{2}}\,
\{\yr(\delta)\}^{1/2}\left(1+\frac{1}{\{\yr(\delta)\}^p}\right)
\,e^{-\yr(\delta)/2}\right] .
\label{eq: F}
\end{eqnarray}
The expression (\ref{eq: pdf_power}) with (\ref{eq: y-function}) and
(\ref{eq: F}) is a main analytical result in this paper. Qualitatively, 
it reveals the strong dependence of the halo density profiles and the 
sensitivity to the initial spectra.

Figure \ref{fig: ff} shows the behavior of the function 
$F(\delta;\alpha,n)$ for various values of $n$ ({\it left panel}) and for 
$\alpha$ ({\it right pane}l). 
Note again that the dependence of the power index $n$ on the PDF
enters only through the function $F(\delta;\alpha,n)$. 
It is clear from both panels that the slope of the function $F$ becomes 
sensitive to both $n$ and $\alpha$ in the range 
$\delta/\delta_*>1$, corresponding to the mass range 
$M\ga M_*$. The sensitivity to the power index $n$ simply results 
from the steep exponential tails of the halo mass function, which becomes 
very sensitive to the spectral index in such mass range. 
And the sensitivity to the slope of the halo profile $\alpha$ is also 
explained from the exponential tails of mass function, that is,  the 
variation in the relation between characteristic smoothed density 
$\delzero$ (eq.[\ref{eq: def_of_delpow}]) and halo mass $M$ due to a small 
change of $\alpha$ is significantly enhanced through such exponential 
dependences. 
On the other hand, the slope of the function $F$, in turn, becomes  
insensitive to both $n$ and $\alpha$ at the region $\delta/\delta_*<1$. 
This is because 
in such range (corresponding to the mass range of $M< M_*$) the mass function
only weakly depends on the spectral index $n$ and scales as  
$n(M)\propto M^{(n-9)/6}$. 
Therefore, within the range $\delta/\delta_*<1$, the PDF is 
almost solely determined by the contribution from the 
halo profile and varies as $\propto \delta^{-(3/\alpha+1)}$, 
while at the region 
$\delta/\delta_*>1$, the function $F$ varies much faster than 
$\delta^{-(3/\alpha+1)}$, 
thus the shape of PDF is determined by the contribution from the 
exponential tails of mass function.

\begin{figure}
  \begin{center}
\begin{minipage}{0.49\textwidth}
    \epsfxsize=8.0cm
    \epsfbox{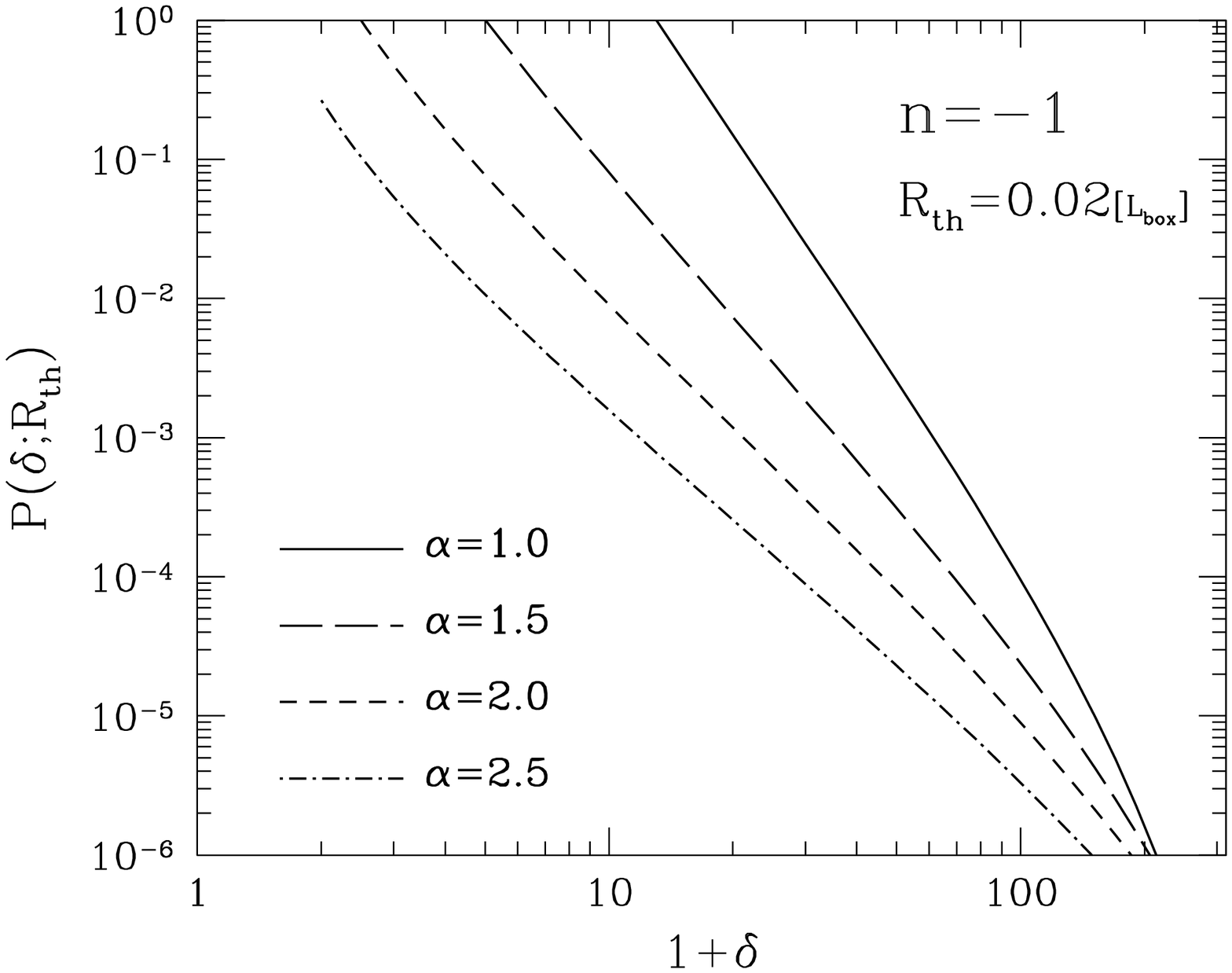}
\end{minipage}
\begin{minipage}{0.49\textwidth}
    \epsfxsize=8.0cm
    \epsfbox{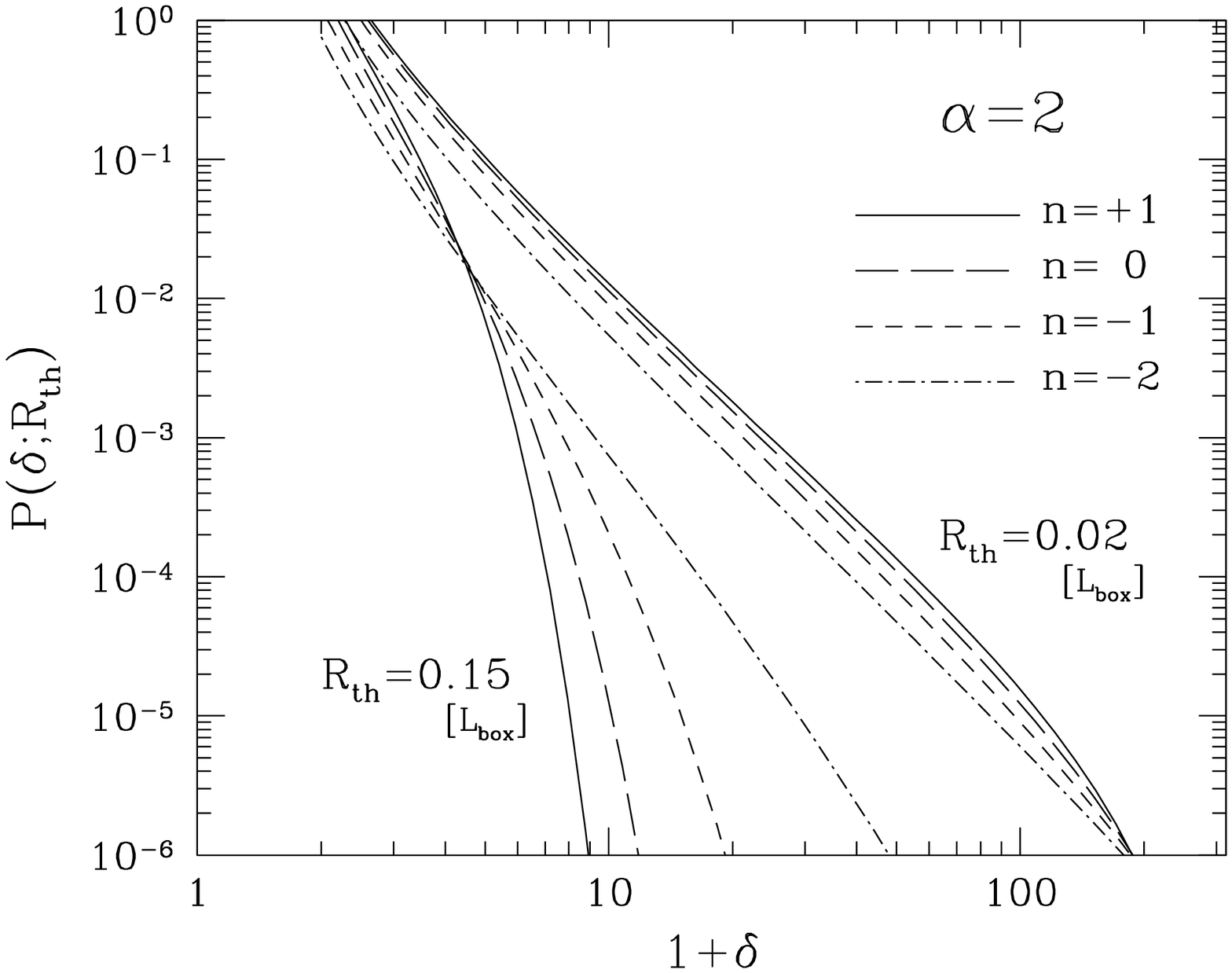}
\end{minipage}
  \end{center}
    \caption{Dependence of the halo density profiles({\it left}) 
      and initial spectra({\it right}) on the approximate expression 
        of the one-point PDF (\ref{eq: pdf_power}) in scale-free models. 
        Here, the normalization of rms mass fluctuation $\sigma_M$ is 
        determined by introducing the auxiliary length scale $L_{\rm box}$ 
        and setting $\sigma_M=1$ at $R_{M}=0.1L_{\rm box}$ in each panel.
        While the initial spectrum and the top-hat smoothing radius are 
        respectively chosen as $n=-1$ and $\rth=0.02L_{\rm box}$ in 
        left panel, the right panel fixes the slope of the power-law 
        profile to $\alpha=2.0$.
    \label{fig: halo_profile} }
\end{figure}

Figure \ref{fig: halo_profile} shows the approximate expression of one-point 
PDF (\ref{eq: pdf_power}) for various slope of the halo profiles ({\it left}) 
and for various initial spectra ({\it right}). For definiteness, 
we assume the Einstein-de Sitter 
universe (density parameter $\Omega_0=1$, cosmological  constant 
$\lambda_0=0$) and the non-linear mass $M_*$ in each model is 
determined by introducing the artificial length scale $L_{\rm box}$ 
and normalizing the rms fluctuation $\sigmam$ to unity at 
$R_{M}=0.1L_{\rm box}$. 
The smoothing radii are specifically chosen as both the 
large $\delta_*$ ($\sim 30-90$) ($\rth=0.02L_{\rm box}$) and the 
small $\delta$ ($\sim 0.3-1.5$) ($\rth=0.15L_{\rm box}$). 
Table {\ref{tab: delta*} summarizes the values of $\delta_*$ 
for the models shown in Figure \ref{fig: halo_profile}.

\begin{table}
\caption{The characteristic smoothed density $\delta_*$ for models presented 
in Figure \ref{fig: halo_profile}.}
\label{tab: delta*}

\begin{center} 
{\large
\begin{tabular}{cccccccc} 
\hline
\multicolumn{2}{c}{Left panel} &&& \multicolumn{4}{c}{Right Panel }\\
\\
\multicolumn{2}{c}{$n=-1$, $\rth=0.02[L_{\rm box}]$} &&
\multicolumn{2}{c}{$\alpha=2$, $\rth=0.02[L_{\rm box}]$} &&
\multicolumn{2}{c}{$\alpha=2$, $\rth=0.15[L_{\rm box}]$}\\
$\alpha$ & $\delta_*$ && $n$ & $\delta_*$ && $n$ & $\delta_*$\\
\hline\hline
1.0 & 93.7 && +1 & 83.3 && +1 & 1.48 \\
1.5 & 68.0 && 0  & 70.0 && 0  & 1.24 \\
2.0 & 49.4 && -1 & 49.4 && -1 & 0.88 \\
2.5 & 35.9 && -2 & 17.4 && -2 & 0.31 \\
\hline
\end{tabular}
}
\end{center}
\end{table}

In left panel, the spectral index are fixed to $n=-1$ but the slope of 
halo profile $\alpha$ varies from $1$ to $2.5$, while in right panel,  
$\alpha$ is fixed to $2$ but $n$ varies from $-2$ to $+1$.
It is clear from the cases of $\rth=0.02L_{\rm box}$ in both 
panels that at the region $\delta\la \delta_*$, the slope of PDF hardly
depends on $n$ but sensitively depends on $\alpha$.
On the other hand, at $\delta\ga \delta_*$ (see the cases of 
$\rth=0.02L_{\rm box}$ in the right panel), the shape of PDF strongly 
depends on $n$.
It follows from these results that if one takes smaller smoothing scale 
than the virial radius of the halo mass $M_*$ (thus $\delta_*\ga100$), 
the slope of the PDF almost solely determined by the slope of the halo 
profile at the intermediate range of $5<\delta<50$, while the shape of PDF 
strongly depends both on $\alpha$ and $n$ 
if we conversely take a larger smoothing scale (so that $\delta_*\la10$).
Figure \ref{fig: pdf_diff_M} shows this point clearly.
In the figure, the PDF in the case of $n=-1$ and $\alpha=2$ 
is shown restricting the contribution of halo mass to some small mass ranges. 
It turns out that at $\delta>\delta_*$, the major contribution to the PDF 
at a certain value of $\delta$ comes from halos whose characteristic
mass scale $\delzero$ roughly coincides with $\delta$, 
thus the shape of PDF reflects shape of the halo mass function. 
At $\delta<\delta_*$, however, halos with $\delta_0=\delta$ 
no longer lead to a major contribution, but 
halos whose characteristic density $\delzero$ is larger than $\delta$ 
instead contribute to the PDF. In this case, PDF reflects the profile 
of the halos. 

\begin{figure}
  \begin{center}
    \epsfxsize=9.0cm
    \epsfbox{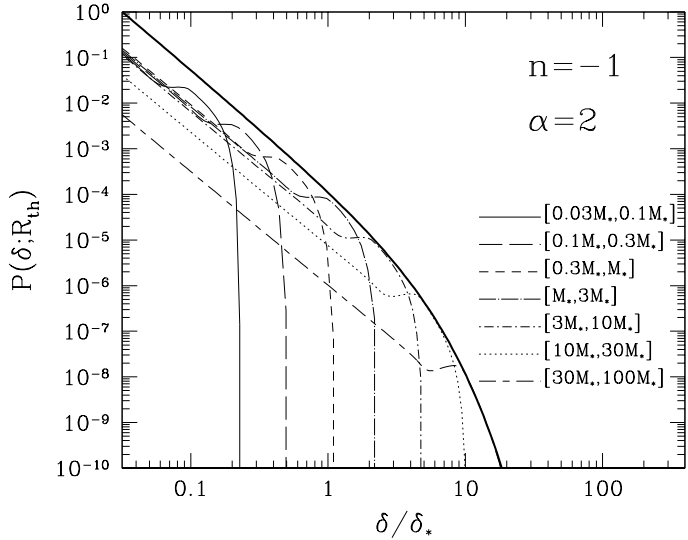}
  \end{center}
    \caption{Contribution to the tails of PDF from different halo mass range 
      for the model with $n=-1$ and $\alpha=2$. The thick solid lines are 
      the exact calculation of equation (\ref{eq: pdf}) with (\ref{eq: cdf}),  
      while the thin lines represent the results restricting the range of 
      integration. Here, the smoothing radius is specifically chosen as 
      $\rth=0.02[L_{\rm box}]$, but the the horizontal scaling  
      $\delta/\delta_*$ is not affected by any specific choice of $\rth$. 
    \label{fig: pdf_diff_M} }
\end{figure}

\section{Comparison with N-body simulations}
\label{sec: comparison}
%
%
%
%
We are now in a position to compare the one-point PDFs 
from the dark halo approach with those obtained from N-body simulations. 
To investigate this issue, the analytic results in the previous section 
turn out to become inadequate because of the non power-law nature of the 
realistic halo profiles. Hence, all the model predictions 
presented below are based on the NFW profile (\ref{eq: NFW_profile}). 
In computing the analytic PDF (\ref{eq: pdf}), the top-hat smoothed halo 
$\rhosm$ is numerically evaluated in the Fourier space, in which we use 
the following fitting function $y(k;M)$ for the Fourier transform of the 
NFW profile $\rhonfw/\overline\rho$ (Ma \& Fry 2000):  
\begin{equation}
y(k;M)\,=\,\frac{4\pi}{3}\delta_{c}\,r_{s}^{3}\,\,
\frac{\ln(e+1/q)-\ln[\ln(e+1/q)]}
        {\left(1+q^{1.1}\right)^{(2/1.1)}}, 
\label{eq: Fourier_NFW}
\end{equation}
with $q=k r_s$.

\begin{table}
\caption{The estimated values of the rms fluctuation amplitude $\sigma(\rth)$ 
        in Figure \ref{fig: cf_simul} from the simulation data 
        (labeled $\sigma_{\rm sim}$, the non-linear fitting formula 
        by Peacock \& Dodds (1996) (labeled by $\sigma_{\rm PD}$) 
and $\delta_*$.}
\label{tab: sigma_table}

\vspace*{0.0cm}

  \begin{center}
{\large
\begin{tabular}{|ccccccccc|} 
\hline
\makebox[1.0cm]{} & \multicolumn{3}{c}{LCDM} & \makebox[1.0cm]{} 
&\makebox[0.5cm]{} & \multicolumn{3}{c}{scale-free($n=-1$)} 
\\
\makebox[1.5cm]{$\rth[h^{-1}$Mpc]} & \makebox[1.0cm]{$\sigma_{\rm sim}$} 
& \makebox[1.0cm]{$\sigma_{\rm PD}$} 
& $\delta_*$
&\makebox[0.5cm]{} &\makebox[1.5cm]{$\rth[L_{\rm box}]$} 
& \makebox[1.0cm]{$\sigma_{\rm sim}$} & \makebox[1.0cm]{$\sigma_{\rm PD}$}
& $\delta_*$
\\ 
\hline\hline
2 & 4.17 & 4.08 & 11.0 && 0.02 & 4.81 & 4.99 & 35.8 \\
6 & 1.37 & 1.40 & 0.569 && 0.05 & 1.80 & 1.95 & 3.25 \\
18& 0.44 & 0.50 & 0.0272 && 0.15 & 0.54 & 0.64 & 0.165 \\ 
\hline
\end{tabular}
}
  \end{center}
\end{table}

For the present purpose, we specifically treat the N-body simulation 
data for the scale-free models with initial power spectra  
$P(k)\propto k^n$ $(n=1,0,-1,-2)$(Jing 1998) and the CDM model 
with cosmological constant(Lambda CDM, hereafter LCDM; Jing \& Suto 1998). 
All the models employ $N=256^{3}$ dark matter particles in a periodic 
comoving cube, $L_{\rm box}^{3}$, where the box size of the LCDM model is 
specifically chosen as $L_{\rm box}=100h^{-1}$Mpc. The gravitational force 
calculation is based on the P$^3$M algorithm. 
While the scale-free models assume an Einstein-de Sitter 
universe, cosmological parameters of the LCDM model are chosen as 
$(\Omega_0, \lambda_0, h, \sigma_8)=(0.3, 0.7, 0.7, 1.0)$, where 
the normalization $\sigma_8$ means the linear rms fluctuation at 
$\rth=8h^{-1}$Mpc. As for the scale-free models, the normalization of 
the density fluctuation is determined by 
setting the linear rms fluctuation to unity at $\rth=0.1L_{\rm box}$.

\begin{figure}
  \begin{center}
\begin{minipage}{0.49\textwidth}
    \epsfxsize=8.5cm
    \epsfbox{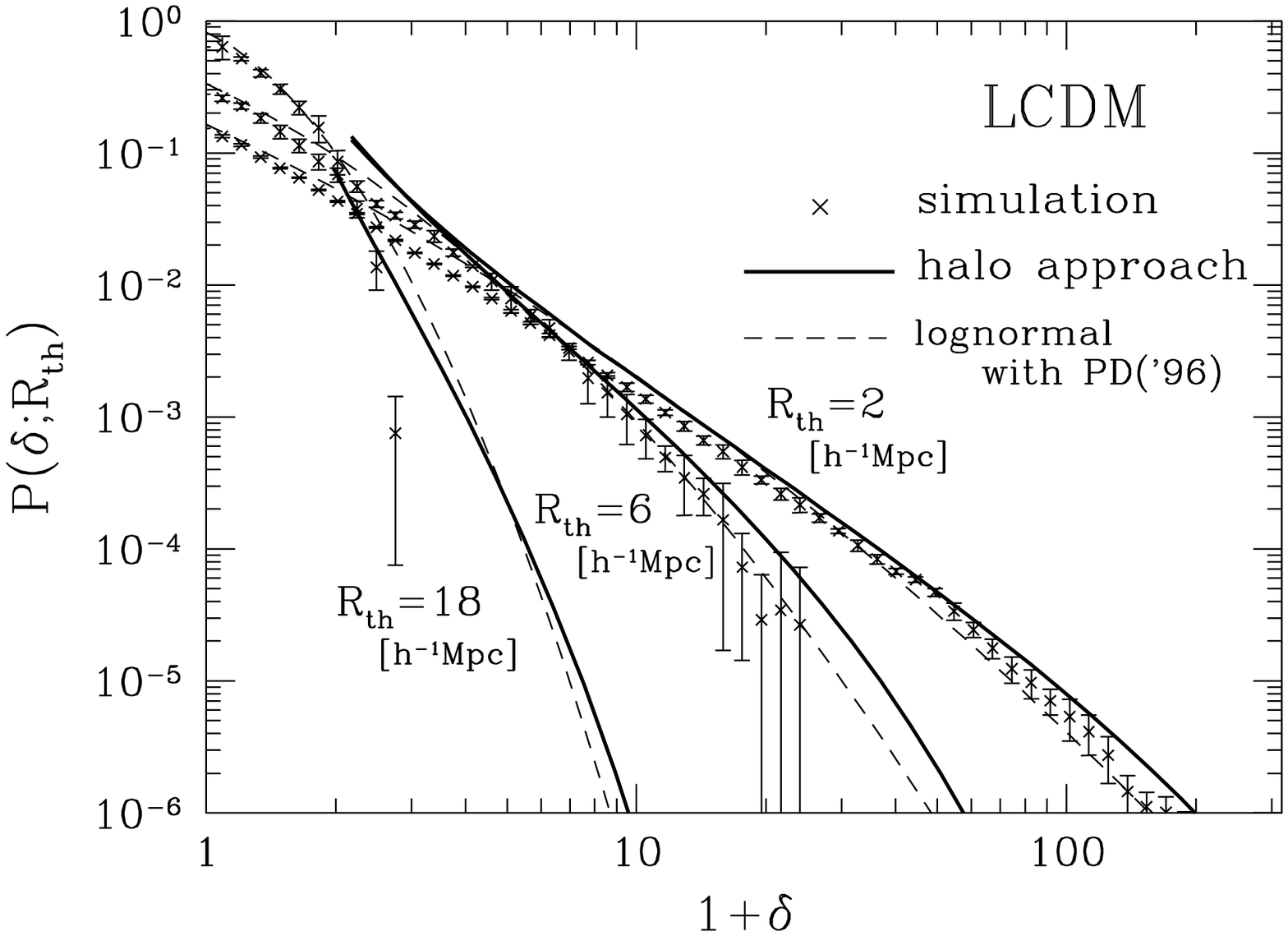}
\end{minipage}
\begin{minipage}{0.49\textwidth}
    \epsfxsize=8.5cm
    \epsfbox{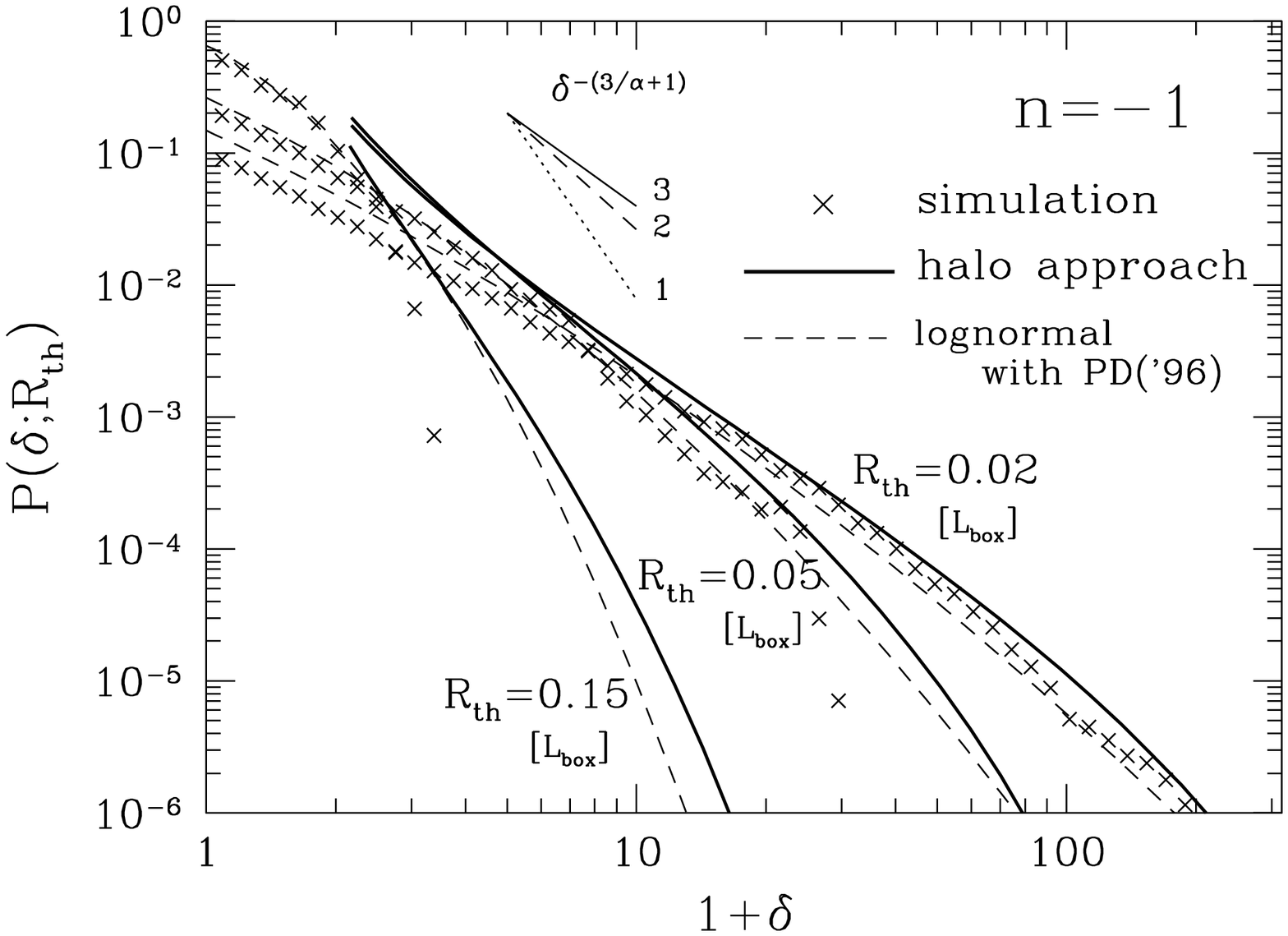}
\end{minipage}
\end{center}
    \caption{Comparison of one-point PDF between analytic models
        ({\it solid, dashed}) and N-body simulations({\it crosses}) 
        in LCDM model({\it left}) and 
        scale-free model with $n=-1$({\it right}). In each panel, 
        the results from dark halo approach are shown in solid lines, 
        while the dashed lines indicate the log-normal model 
        prediction (\ref{eq: log-normal_pdf}). 
        In plotting the log-normal PDF, 
        the rms fluctuation of density field $\sigma$ is evaluated 
        using the fitting formula by Peacock \& Dodds (1996). The error-bars 
        in left panel denote the 1-$\sigma$ errors  
        among three different realization of simulation data. Also, 
        in the right panel, the 
        asymptotic slopes of the dependence $\delta^{-(\alpha/3+1)}$ 
        with $\alpha=1$, $2$, and $3$ are shown. 
            \label{fig: cf_simul}  }
\end{figure}

Figure \ref{fig: cf_simul} shows the N-body results 
of the one-point PDF for the top-hat smoothed density field in various 
smoothing radii ({\it crosses}). The left panel shows the results in LCDM 
model, 
while the right panel plots the PDFs in the scale-free model with the 
initial spectrum $n=-1$. The error-bars in the left panel indicate 
the 1-$\sigma$ errors among three different realizations. 
The smoothing radii in left(right) panel are chosen 
as $\rth=2, 6, 18h^{-1}$Mpc ($\rth=0.02,0.05,0.15L_{\rm box}$),which are 
typically scaled from non-linear to weakly non-linear regime. 
In Table \ref{tab: sigma_table}, computed values of the rms amplitude 
$\sigma$ are presented(labeled by $\sigma_{\rm sim}$). 
Obviously, the degree of non-linearity becomes 
significant as decreasing $\rth$ and the resultant PDFs 
show much longer non-Gaussian tails.

In Figure \ref{fig: cf_simul},  the one-point PDFs from the dark halo approach 
are shown in solid lines. Within the validity range $\delta\simgt1$,  
in both panels, the dark halo approach reproduces the non-Gaussian tails of 
simulated PDF reasonably well. The predictions at large smoothing 
radius $\rth=18h^{-1}$Mpc($0.15L_{\rm box}$) somewhat over-predict the 
simulations, however, as have been pointed out by Kayo et al.(2001), 
this is most likely due to the finite volume effect of the 
N-body simulations.  

For comparison, the dashed lines in Figure \ref{fig: cf_simul} 
show the empirical log-normal distribution. The one-point PDF 
of the log-normal model is analytically expressed as 
\begin{equation}
P_{\rm LN}(\delta)=\frac{1}{\sqrt{2\pi\,\sigma_{\rm LN}^2}}\,\,
\exp
\left\{-\frac{[\log(1+\delta)+\sigma_{\rm LN}^2/2]^2}{2\sigma_{\rm LN}^2}\right\}
\,\,\frac{1}{1+\delta},   
\label{eq: log-normal_pdf}
\end{equation}
where the single parameter $\sigma_{\rm LN}$ 
is related to the rms amplitude of density fluctuation at a given 
smoothing radius $\rth$, $\sigma(\rth)$: 
\begin{equation}
\sigma_{\rm LN}^2 = \log[1+\sigma^2(\rth)].
\end{equation}
For a given set of cosmological parameters, the quantity $\sigma(\rth)$ 
is computed using a fitting formula of non-linear power spectrum 
by Peacock \& Dodds (1996) or even by the dark halo approach 
(Seljak 2000; Ma \& Fry 2000; Scoccimarro et al. 2001; Hamana, Yoshida \& Suto 2001). 
The resultant values are summarized 
in Table \ref{tab: sigma_table} (labeled by $\sigma_{\rm PD}$) 
and adopting those, the log-normal predictions 
(\ref{eq: log-normal_pdf}) are plotted. In this sense, 
both empirical and analytic models in Figure \ref{fig: cf_simul} provide a 
fairly self-consistent prediction without invoking any information of 
simulation data.  

As mentioned in Kayo et al. (2001), the rms amplitude 
$\sigma_{\rm PD}$ takes a slightly larger value than that of the simulation, 
$\sigma_{\rm sim}$, which should be ascribed to the finite box-size of the 
simulations. Apart from this small discrepancy, the log-normal PDFs approximate 
the simulated PDF fairly accurately in both low-density and high-density 
regions. The most striking evidence is that the tails of analytic PDF almost 
resembles those of the log-normal PDF, which is also consistent with the 
prediction from perturbation theory(Bernardeau 1994b; Bernardeau \& Kofman 
1995). According to the Table \ref{tab: sigma_table}, the theoretical 
estimate of characteristic smoothed density $\delta_*$ roughly varies 
from $\delta_*\sim 0.03$ to $40$ as decreasing the smoothing radius. Thus, 
the tails of PDF at smoothing radius $\rth=2h^{-1}$Mpc or $0.02L_{\rm box}$ 
basically reflect the shape of the halo profiles and they scale as 
$\delta^{-(\alpha/3+1)}$. Comparing the shape of PDFs with 
the slope of the dependence $\delta^{-(\alpha/3+1)}$ superimposed in   
the right panel of Figure \ref{fig: cf_simul}, 
the non-Gaussian tails at smaller smoothing radii 
can be regarded as a consequence of outer slope of the NFW profile, 
$\alpha=3$. 

\begin{figure}
  \begin{center}
\begin{minipage}{0.49\textwidth}
    \epsfxsize=8.5cm
    \epsfbox{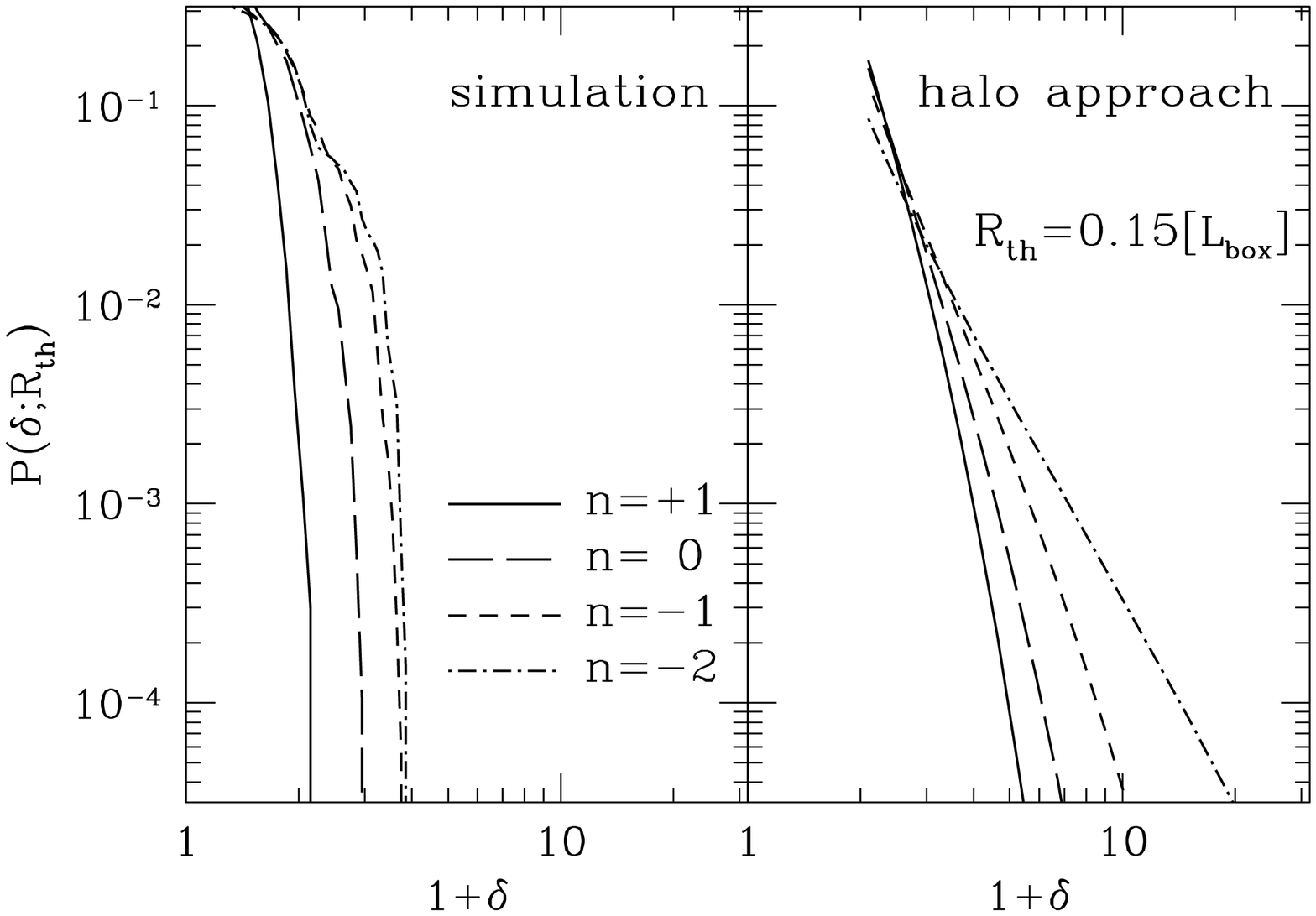}
\end{minipage}
\begin{minipage}{0.49\textwidth}
    \epsfxsize=8.5cm
    \epsfbox{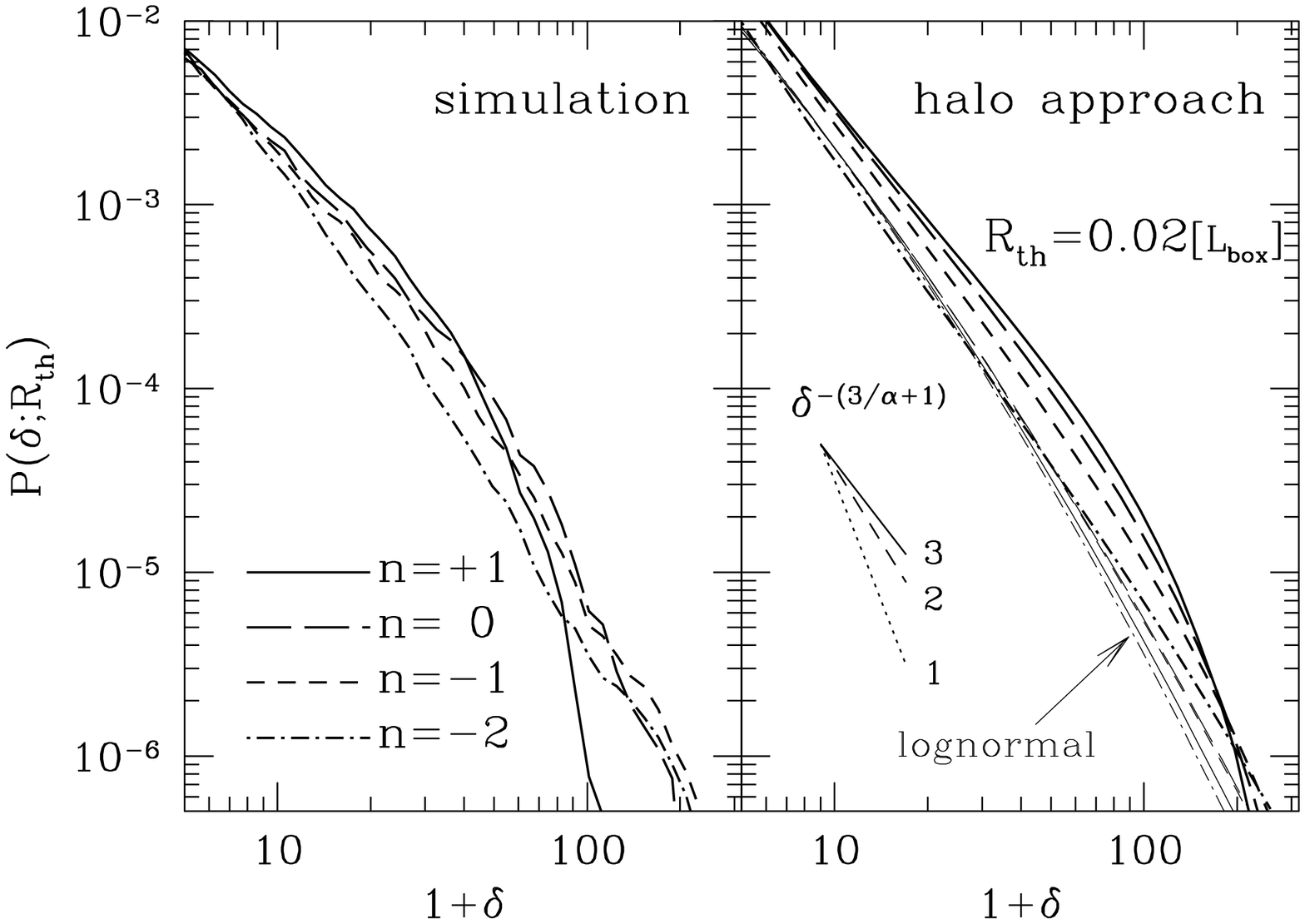}
\end{minipage}
\end{center}
  \caption{Comparison of non-Gaussian tails of PDF between 
    dark halo approach and N-body simulations in cases with scale-free 
    initial spectra. The left(right) panel depicts the results at 
    the smoothing radius $\rth=0.02L_{\rm box}$($\rth=0.15L_{\rm box}$). 
    For comparison, we also plot the log-normal PDFs 
    in the right-side of the right panel ({\it thin lines}), 
        where rms fluctuation of log-normal PDF, $\sigma$ is 
        calculated from simulations. Also, the asymptotic slope of 
        the dependence $\delta^{-(\alpha/3+1)}$ with $\alpha=1$, $2$, and $3$ 
        are shown. 
    \label{fig: cf_init_spectra}}
\end{figure}

Figure \ref{fig: cf_init_spectra} shows weak and strong 
sensitivity of the non-Gaussian tails to the initial spectrum in 
scale-free models. Specifically choosing the smoothing radius as 
$\rth=0.15L_{\rm box}$({\it left}) and $\rth=0.02L_{\rm box}$({\it right}), 
the results from the dark halo approach are plotted in right-side 
of each panel, while the left-side shows the simulation results. 
For comparison, the log-normal PDFs are also plotted 
in right-side of the right panel({\it thin-lines}), 
where rms amplitude $\sigma$ is calculated from simulations.   
In cases with large smoothing radius $\rth=0.15L_{\rm box}$, 
the simulated PDFs rapidly fall off and they eventually vanish,  
while the analytic model of PDF shows a slightly longer non-Gaussian 
tail. This discrepancy is basically ascribed to the finite volume size of 
the simulations (Kayo et al. 2001), however, the tendency seen in both 
panels is qualitatively similar. The analytic model 
reproduces the systematic behavior for the variation of spectral index $n$ 
(see also the right-panel of Fig.\ref{fig: halo_profile}).  
On the other hand, for smaller smoothing radius with $\rth=0.02L_{\rm box}$, 
except for the noisy data of scale-free model with $n=+1$,  
both left- and right-side of the panel show not only qualitatively but also 
quantitatively similar dependence on the initial power spectra. 
Again, the asymptotic behavior of the PDF is quite similar to the results 
with steeper power-law profile $\alpha\sim 3$ (compare with the slope 
$\delta^{-(\alpha/3+1)}$ superimposed in the figure). 
As a consequence, the non-Gaussian 
tails of PDF seem to resemble the log-normal PDF insensitive to the initial 
spectrum, as has been reported by Kayo et al. (2001). 
At a closer look at tails of PDF, however, the simulated PDFs  
still exhibit the initial spectrum dependence and they 
slightly deviate from log-normal PDFs ({\it thin-lines}). 
In contrast, the analytic PDF from the dark halo approach 
successfully reproduces this weak dependence. 
Note that the theoretical estimate of the characteristic smoothed density 
yields 
$\delta_*\sim 10$--$70$ at $\rth=0.02L_{\rm box}$, which suggests that the 
the simulated PDF will show a strong sensitivity to the spectral index 
at the high density tail $\delta/\delta_*\gg1$, and thereby 
a large discrepancy between log-normal prediction and simulation is expected. 
Interestingly, there is some indication at the region  
$\delta\simgt 200$ in both simulation and halo approach, where the 
systematic variation for $n$ apparently ceases.

Therefore, we conclude that the analytic model based on the dark halo 
approach reasonably predicts the general tendency in both weakly 
and strongly non-linear regime and it 
quantitatively provides a useful approximation to the 
non-Gaussian tails of simulated PDF in the strongly non-linear regime. 
%
%
%
%
%
%
%
%
%
%
%
%
%
%
%
%
%
%
%
%
%
\section{Discussion \& Conclusion}
\label{sec: conclusion}
%
%
%
%
%
%
%
In this paper, utilizing the dark halo approach, we have presented 
a simple analytic model characterizing the non-Gaussian tails of 
one-point PDF and investigated the sensitivity to both the initial power 
spectrum and the halo density profiles in the non-linear regime. 
Assuming the power-law profile of dark halos and the scale-free 
initial spectrum, the approximate expression for the tails of PDF is 
analytically derived (eq.[\ref{eq: pdf_power}] with [\ref{eq: y-function}]
[\ref{eq: F}]). Using this result, we explained the 
sensitivity to the initial spectrum seen in the simulation and 
the significant dependences on the halo profiles. Further, quantitative 
comparison with N-body simulation was made, which leads to 
the conclusion that the analytic model reasonably reproduces the 
simulated PDF qualitatively and quantitatively in a 
strongly non-linear regime. 
Then, the tails of simulated PDF 
turn out to resemble the log-normal distribution (\ref{eq: log-normal_pdf}) 
at the region $\delta\simlt 100$ in a strongly non-linear regime.

In our model, the key quantity is the characteristic smoothed 
density $\delta_*=\rhosm(0;R,M_*)/\overline\rho$, 
which discriminates the weak/strong sensitivity to the 
halo profile and initial power spectrum (see eq.[\ref{eq: y-function}]). 
At $\delta/\delta_*\simlt1$, 
the tails of PDF basically follow the halo profile and varies as 
$\delta^{-(3/\alpha+1)}$, while 
the PDF at $\delta/\delta_*\simgt1$ exhibits a strong sensitivity to 
the initial spectrum, which can be ascribed to the steep exponential 
tails of halo mass function. In CDM models at present, 
the characteristic smoothed density typically takes the value 
$\delta_*\gg 1$ for a small smoothing scale $\rth\sim 1$Mpc$/h$ and 
conversely $\delta_*\ll 1$ for a large smoothing scale $\rth > 10$Mpc$/h$. 
Therefore, the log-normal behavior seen in the 
non-linear regime can be regarded as a consequence of the universality of 
the halo density profile. Further considering the fact that the slope of 
the simulated PDF is similar to that of the analytic results with 
$\alpha\simeq 3$, the weak sensitivity to the initial spectrum is 
especially due to the universal outer profile of 
$\rho \propto r^{-3}$.

The qualitative feature in the dark matter PDFs  
might be also the case for the PDF of galaxies,  
which has been known to be well approximated by the log-normal PDF (e.g., 
Hamilton 1985; Bouchet et al. 1993; Kofman et al. 1994) 
as the density PDF is. 
The number density of galaxies within the clusters is known to be 
fitted well by the King model (King 1962), which has a radial profile of 
$\rho_g \propto [1+(r/r_c)^2]^{-3/2}$, where $r_c$ is a core radius of 
order 0.1$h^{-1}$Mpc (Wu \& Hammer 1993 and references therein). 
Thus, on suitably larger smoothing scale $\rth\simgt r_c$, smoothed density 
profile becomes $\rho_g \propto r^{-3}$, at an outer part. 
Therefore, it is very likely that the similarity in the shape of the 
PDF tails between the dark matter and the galaxy is due to simply their 
similar distribution within halos.

Further, the good agreement in the non-Gaussian tails of density PDF between 
the halo approach prediction and N-body data indicates that the halo model 
is not only used to qualitative study but also allows us to make a 
quantitative prediction. 
Although the dark matter PDF is hard to measure, it may be applied to 
related measurements such like the weak lensing convergence which is a 
line-of-sight projection of the matter density weighted by the lensing 
efficiency and is obtained from a coherent distortion in distant 
galaxy images (see Mellier 1999; Bartelmann \& Schneider 2000 for reviews). 
Taruya et al. (2002) showed that the lensing convergence PDF is 
well approximated by the log-normal model and pointed out that 
the lensing convergence field can be regarded as a simple projection 
of the density field. 
This implies that one can safely apply the halo approach to the lensing 
convergence PDF. 
Actually, Kruse \& Schneider (2000) employed the halo approach to predict the 
non-Gaussian tail in PDF of lensing convergence arguing it may 
be used to discriminate the cosmological model.
Their prediction was tested against numerical experiments and a 
good agreement was found (Reblinsky et al. 1999). 
Since the approach adopted by Kruse \& Schneider (2000) is essentially the 
same as described in this paper, one naturally expects that the tails 
of lensing convergence PDF basically follows the outer slope of the halo 
profiles at smaller angular scale $\theta\sim1'$-$5'$, and the log-normal 
behavior found by Taruya et al. (2002) can be explained by similar reason 
to the mass density PDF.

Therefore, as far as the CDM model with a Gaussian initial condition is 
concerned, theoretical prediction based on the dark halo approach consistently 
explains the log-normal nature of observable clustering statistics  
as a common feature in the hierarchical clustering universe. 
This conversely indicates that the statistics related to the non-Gaussian 
tails of PDF provide a powerful cosmological probe to the nature of 
dark matter clustering, which enables us to constrain the number 
density and the density profile of dark halos. 
Probably, one promising application is the weak lensing statistics, 
as has been demonstrated by Kruse \& Schneider (2000). Recently, from 
the $2.1$ deg$^2$ image taken with a wide field camera on the prime focus 
of Subaru Telescope, non-Gaussian signature of the lensing convergence was 
detected evaluating the peak distribution function (Miyazaki et al. 2002). 
The number count of high positive peaks, which are thought to come from the 
dark halos, was compared with the theoretical prediction using halo approach 
and showed a consistent result with NFW profiles under the LCDM cosmology. 
Note, however, that the analytic models of PDF from the dark halo approach 
currently limit the applicability to highly non-Gaussian tails in a strongly 
nonlinear regime. To obtain a more stringent constraint on the cosmological 
parameter and/or the number density of dark halos, we should further improve 
the applicability to the low-density region up to the observable scale of 
the non-linear regime. In the light of this, the perturbative construction 
of PDF by Bernardeau \& Valageas (2000) or the non-perturbative modeling 
from the hierarchical ansatz (Munshi \& Jain 2000; Valageas 2000), as well 
as the empirical log-normal model (Taruya et al.2002) could be useful and 
be regarded as a complementary approach. Combining the present work with 
these approaches would be possible to improve the prediction dramatically 
and will be discussed elsewhere. 
%
%
%
%
%
%
%
%
%
%
%
%
\section*{Acknowledgments}
%
%
%
We thank Y.P.Jing for kindly providing us his N-body data,   
Y.Suto for reading of the manuscript and 
critical comments, M.Takada and K.Yoshikawa for valuable discussion. 
I.K is supported by Takenaka-Ikueikai Fellowship. T.H acknowledges 
supports from 
Japan Society for Promotion of Science (JSPS) Research Fellowships. 
A.T is supported by the grand-in-aid for Scientific Research 
of JSPS (No.$1470157$). 
%
%
%
%
%
%
%
%
%
\section*{Appendix A: Smoothed density profile for power-law halos}
\label{appendix: smoothed_profile}
%
%
%
%
%
%
In this appendix, adopting the top-hat filter function 
(\ref{eq: top-hat_filter}), we derive the exact expression for the smoothed 
halo profiles. Substituting the definition (\ref{eq: top-hat_filter}) into 
(\ref{eq: rho_smooth}), the expressions for smoothed density profile can be 
reduced to  
\begin{eqnarray}
\rhosm(r;\rth,M) &=& \frac{3}{4\pi \rth^3}\,\,\int d^3\bfr' 
        \,\,\Theta(\rth-|\bfr'|)\,\,\rho(|\bfr-\bfr'|;M)
\nonumber\\
&=& \frac{3}{2\rth^3}\,\,\int_0^{\rth}ds\, s^2
        \int_{-1}^{1}dx\,\,\rho(\sqrt{r^2+s^2-2rsx\,\,};\,M).
\label{eq: smooth_halo_TH}
\end{eqnarray}
For the power-law profiles (\ref{eq: power-law_profile}), 
all the integrals in the expression (\ref{eq: smooth_halo_TH}) are 
analytically evaluated, which lead to 
\begin{eqnarray}
\rhosm(r;\rth,M) &=& \frac{3\delta_c\,\overline\rho\,r_s^{\alpha}}{2\rth^3}\,
\int_0^{\rth}ds\,s^2\,
\frac{|r-s|^{2-\alpha}-(r+s)^{2-\alpha}}{(\alpha-2)r\,s}
\nonumber\\
&=& \frac{3}{2(\alpha-2)}\,\rhopower(\rth;M)\,\,
\nonumber\\
&&~~~\times \,\left\{
\begin{array}{lcc}
\frac{1}{3-\alpha}\left\{(1-u)^{3-\alpha}+(1+u)^{3-\alpha}\right\}+
\frac{u^{-1}}{4-\alpha}\left\{(1-u)^{4-\alpha}-(1+u)^{4-\alpha}\right\}
& ; & (0< r \leq \rth )
\\
\\
\frac{1}{3-\alpha}\left\{(1+u)^{3-\alpha}-(u-1)^{3-\alpha}\right\}+
\frac{u^{-1}}{4-\alpha}\left\{(u-1)^{4-\alpha}-(1+u)^{4-\alpha}\right\}
& ; & (\rth < r)
\end{array}
\right.
\label{eq: smooth_halo_1}
\end{eqnarray}
Here, the variable $u$ denotes 
$r/\rth$. Notice that the above expressions are valid for the profiles 
with $\alpha\neq2$. For the power-law profile with $\alpha=2$ 
(singular isothermal sphere), the logarithmic dependence of the radius $r$ 
appears, which can be separately evaluated as follows:
\begin{eqnarray}
\rhosm(r;\rth,M) = \frac{3\delta_c\,\overline\rho\,r_s^2}{2\rth^3}\,
\int_0^{\rth}ds\,\frac{s}{r}\,\log\left|\frac{r+s}{r-s}\right|
= \frac{3}{2}\,\rho(\rth;M)\,\left\{1+\frac{1}{2}(u^{-1}-u)\,\,
        \log\left|\frac{u+1}{u-1}\right|\right\}.
\label{eq: smooth_halo_2}
\end{eqnarray}
%
%
While the above results are exact, it is convenient and useful to 
derive the simple analytic expressions for the one-point PDF 
$P(\delta;\rth)$ (see Sec.\ref{sec: results}). 
From the asymptotic behavior of the smoothed halo profiles, we 
approximate the expressions (\ref{eq: smooth_halo_1}) and 
(\ref{eq: smooth_halo_2}) as follows:
\begin{equation}
\rhosm(r;\rth,M)\simeq
\left\{
\begin{array}{lcr}
\left(\frac{3}{3-\alpha}\right)\,\,\rhopower(\rth;M) & ; & 
        r<\left(\frac{3-\alpha}{3}\right)^{1/\alpha}\,\rth
\\
\\
\left(\frac{r}{\rth}\right)^{-\alpha} \rhopower(\rth;M) & ; & 
        r\geq\left(\frac{3-\alpha}{3}\right)^{1/\alpha}\,\rth
\end{array}
\right.
\label{appen_A: approx}
\end{equation}
%
%
The left panel of Figure \ref{fig: rho_smooth_power} compares the 
approximation (\ref{appen_A: approx}) with the exact results 
(\ref{eq: smooth_halo_1}) and (\ref{eq: smooth_halo_2}). 
Apart from the intermediate scales $r/\rth\sim1$, the approximation 
recovers the exact results reasonably well. On the other hand, 
deviation at the intermediate scale slightly increases as increasing 
the slope of profile $\alpha$, which leads to the under-prediction of 
volume factor $V(>\delta;\rth,M)$. Hence, for steeper halo profile 
$\alpha \sim 3$, the analytic PDF using (\ref{appen_A: approx}) 
tends to under-predict the amplitude of PDF as shown in the right panel.

To check the validity of the approximation (\ref{eq: pdf_power}), 
together with the analytic PDF in the left panel of Figure 
\ref{fig: halo_profile}, 
the PDF computed from the exact results without using 
approximation (\ref{eq: rho_power_approx}) is plotted and the results are 
presented in the right panel of Figure \ref{fig: rho_smooth_power} 
({\it thin lines}). As is expected from the left panel, 
the approximate expression (\ref{eq: pdf_power}) for steeper 
halo profiles with $\alpha=2.5$ tends to lower its amplitude, 
but the qualitative behavior still remains correct. 

\begin{figure}
  \begin{center}
\begin{minipage}{0.49\textwidth}
    \epsfxsize=8.0cm
    \epsfbox{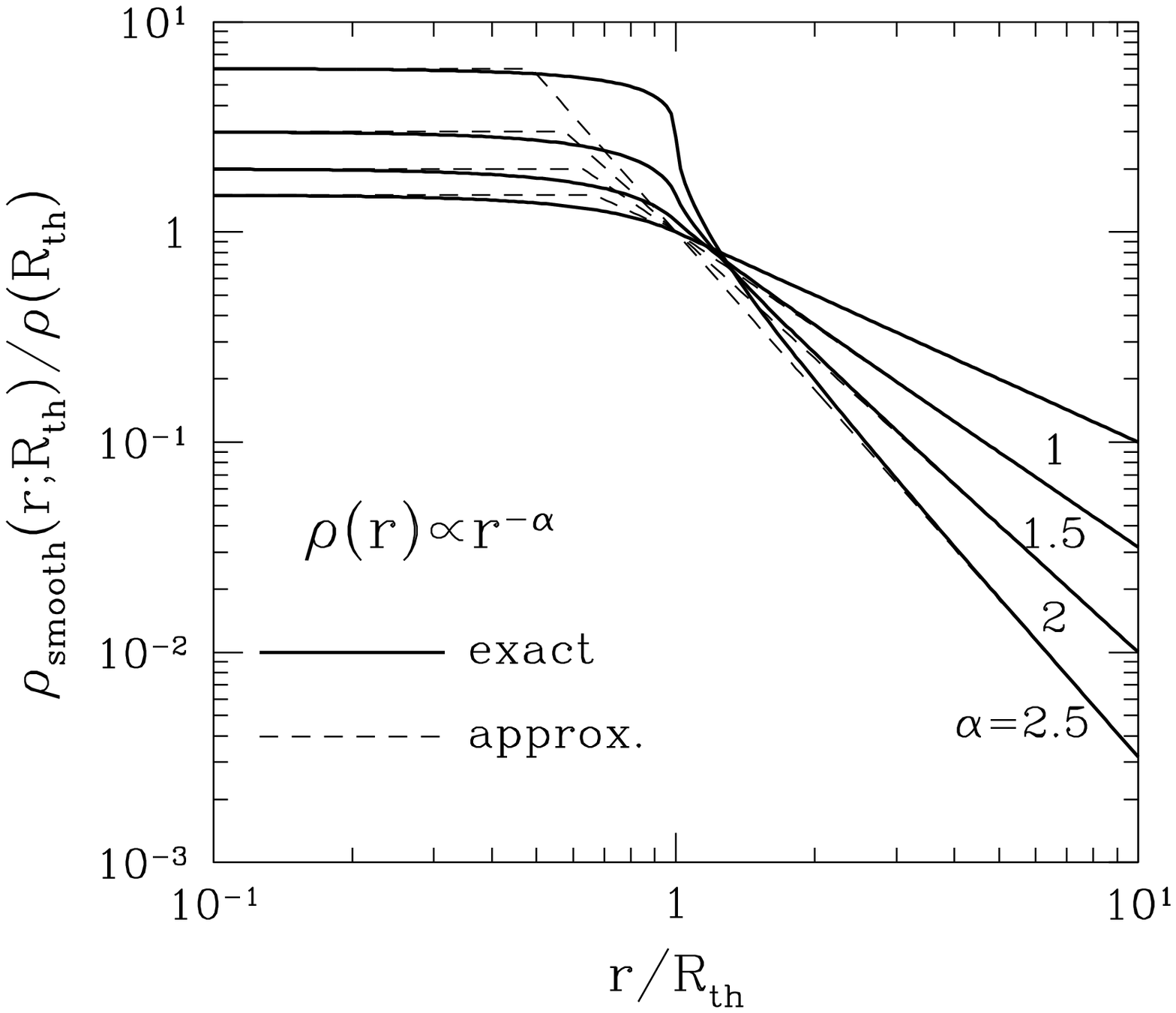}
\end{minipage}
\begin{minipage}{0.49\textwidth}
    \epsfxsize=8.7cm
    \epsfbox{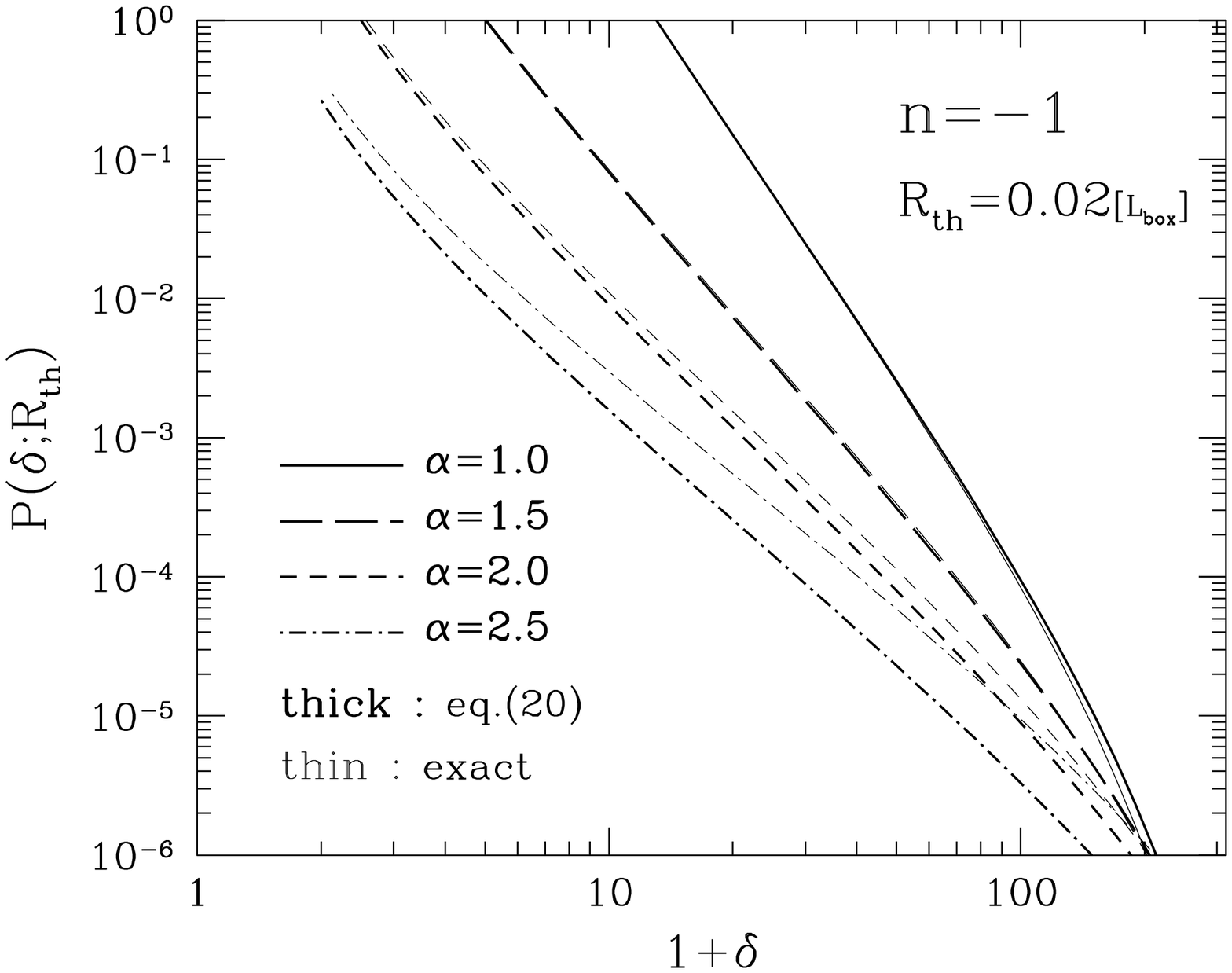}
\end{minipage}
\end{center}
\caption{Left panel: comparison of smoothed power-law profiles between 
exact results 
(Eqs.[\ref{eq: smooth_halo_1}][\ref{eq: smooth_halo_2}]; {\it solid}) 
and approximations (Eq:[\ref{appen_A: approx}]; {\it dashed}).
Right panel: the PDFs computed from the exact results without using the 
approximation (\ref{eq: rho_power_approx}) ({\it thin lines}) 
compared with the approximation results. 
\label{fig: rho_smooth_power}}
\end{figure}

\section*{Appendix B: Effects on uncertainty of the model prediction}
\label{appendix: uncertainty}
%
%
%
%
%
The analytic model of one-point PDF presented in section 
\ref{sec: halo_approach} suffers from uncertainties in the model 
assumptions (mass function of dark halos and halo density profile). 
Among these, the concentration parameter of the NFW halo profiles, 
$c(M)$ might be crucial in comparing the analytic prediction 
with simulations (Sec.\ref{sec: results}), since the sensitivity to 
$c(M)$ has been reported in previous study using dark halo approach 
(e.g., Seljak 2000). Although we adopt the 
fitting form of the concentration parameter by Bullock et al. (20001) 
with a particular set of the parameters $(c_*,\beta)$ (see eq.
[\ref{eq: c(M)}] below), we examine the influence of 
varying the parameters $(c_*,\beta)$. 

\begin{figure}
  \begin{center}
    \epsfxsize=9.0cm
    \epsfbox{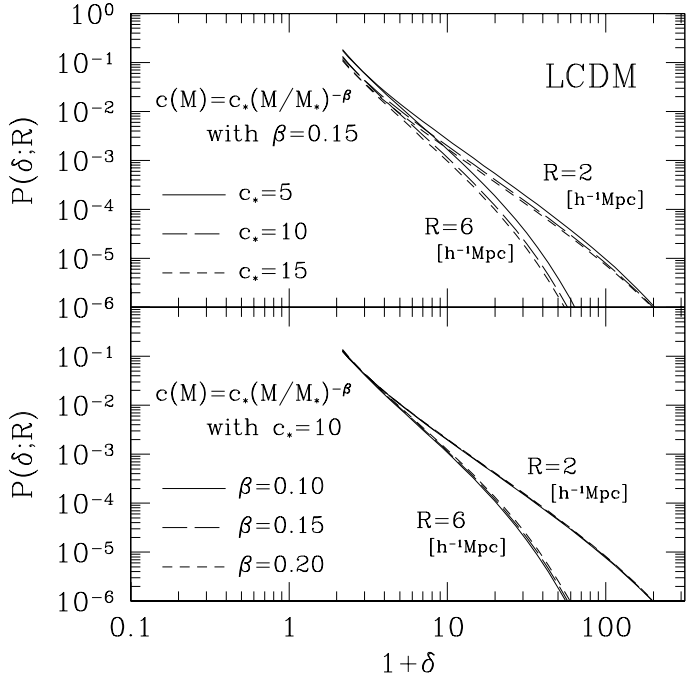}
  \end{center}
    \caption{Influence of varying the concentration parameter $c(M)$ 
        on the analytic model of mass density PDFs for $\rth=2h^{-1}$Mpc 
        case and $\rth=6h^{-1}$Mpc case in LCDM model. 
        The upper(lower) panels show the dependence of 
        concentration parameters fixing the  
        $c_*$($\beta$).         }
    \label{fig: pdf_cM}
\end{figure}

Figure \ref{fig: pdf_cM} shows the dependence of the non-Gaussian tails 
on the variation in concentration parameter assuming the LCDM universe. 
Upper(lower) panels represent the results fixing $\beta$($c_*$). 
Although the influence of the variation $(c_*,\beta)$ turns out to be small, 
the systematic behavior in the tails of PDF seems to vary in a opposite 
sense: decreasing the parameter $\beta$ or increasing $c_*$ 
suppresses the non-Gaussian tails.

The small influence on the tails of PDF can be easily deduced from 
the Fourier transform of the NFW profile (\ref{eq: Fourier_NFW}),  
in which the concentration parameter appears only through 
the factor $\delta_c r_s^3$, leading to the logarithmic dependence of $c(M)$. 
On the other hand, apparently opposite behavior for the variation of 
$(c_*,\beta)$ is explained as follows. First note that decrease of $\beta$ or 
increase of $c_*$ yields more concentrated density field for halos more 
massive than the nonlinear mass $M_*$. This implies that while the 
number of halos in which the central density exceeding a specific value of $\delta$ 
increases, the volume of each density profile exceeding $\delta$ 
conversely decreases in the massive halos. Since the contribution of 
massive halo is relatively dominant in the tails of PDF, the latter effect 
finally wins and accordingly the amplitude of one-point PDF decreases.

Except for the details, the systematic uncertainty of the 
concentration parameter does not affect the final conclusions in section 
\ref{sec: comparison} and the dispersion of $0.2$ in $\ln c$ 
(Jing 2000; Bullock et al. 2001) can be safely neglected in our calculation. 
Note, however, that in the case of the lensing convergence PDF, 
the dependence of $(c_*,\beta)$ becomes slightly larger than that in the 
mass density PDF and the systematic behavior of non-Gaussian tails 
appears in a opposite sense to $P(\delta)$. 
These differences simply result from the projection of the 
three-dimensional density field to the two-dimensional field 
(e.g., Takada \& Jain 2002). 
%
%
%
%
%
%
%
%
%
%
%
%
%


\label{lastpage}

\begin{thebibliography}{99}
\bibitem{BS00} 
Bartelmann M., Schneider P., 2001, Physics Report, 340, 291
\bibitem[\protect\citeauthoryear{Bernardeau}{1994a}]{B1994a} 
Bernardeau F., 1994a, ApJ, 433, 1
\bibitem[\protect\citeauthoryear{Bernardeau}{1994b}]{B1994b} 
Bernardeau F., 1994b, A\&A, 291, 697
\bibitem[\protect\citeauthoryear{Bernardeau \& Kofman}{1995}]{BK1995} 
Bernardeau F., Kofman L., 1995, ApJ, 443, 479
\bibitem[\protect\citeauthoryear{Bernardeau \& Valageas}{2000}]{BV2000} 
Bernardeau F., Valageas P., 2000, A\&A, 369, 1
\bibitem[\protect\citeauthoryear{Bouchet, Strauss, Davis, Fisher, Yahil 
\& Huchra}{Bouchet et al.}{1993}]{BSDFYH1993} 
Bouchet F., Strauss M.A., Davis M., Fisher K.B., Yahil A., \&
 Huchra J.P., 1993, ApJ, 417, 36
\bibitem[\protect\citeauthoryear{Bullock, Kolatt, Sigad, Somerville, 
Kravtsov, Klypin, Primack \& Dekel}{Bullock et al.}{2001}]{BKSSKKPDS2001} 
Bullock J. S., Kolatt T. S., Sigad Y., Somerville R. S., Kravtsov A. V., 
Klypin A. A., Primack J. R., Dekel A., 2001, MNRAS, 321, 559
\bibitem[\protect\citeauthoryear{Coles \& Jones}{1991}]{CJ1991} 
Coles P., Jones B., 1991, MNRAS, 248, 1
\bibitem[\protect\citeauthoryear{Fukushige \& Makino}{2001a}]{FM2001a} 
Fukushige T., Makino, J., 2001a, ApJ, 557, 533 
\bibitem[\protect\citeauthoryear{Fukushige \& Makino}{2001b}]{FM2001b} 
Fukushige T., Makino, J., 2001b, ApJL, submitted (astro-ph/0108014) 
\bibitem[\protect\citeauthoryear{Hamana, Yoshida \& Suto}{Hamana et al.}
{2002}]{HYS2002} 
Hamana T., Yoshida N., Suto Y., 2002, ApJ, 568, 455
\bibitem[\protect\citeauthoryear{Hamilton}{1985}]{H1985} 
Hamilton A.J.S., 1985, ApJ, 292, L35
\bibitem[\protect\citeauthoryear{Jing}{1998}]{J1998} 
Jing Y. P., 1998, ApJ, 503, L9
\bibitem[\protect\citeauthoryear{Jing}{2000}]{J2000} 
Jing Y. P., 2000, ApJ, 535, 30
\bibitem[\protect\citeauthoryear{Jing \& Suto}{1998}]{JS1998} 
Jing Y. P., Suto Y., 1998, ApJ, 494, L5
\bibitem[\protect\citeauthoryear{Jing \& Suto}{2000}]{JS2000} 
Jing Y. P., Suto Y., 2000, ApJ, 529, L69
\bibitem[\protect\citeauthoryear{Juszkiewicz, Bouchet \& Colombi}{1993}]{JBC1993} 
Juszkiewicz R., Bouchet, F., Colombi., 1993, ApJ, 419, L9
\bibitem[\protect\citeauthoryear{Kayo, Taruya \& Suto}{Kayo et al.}{2001}]{KTS2001} 
Kayo  I., Taruya  A., Suto  Y., 2001, ApJ, 561, 22
\bibitem{K62}King I. R., 1962, AJ, 67, 471
\bibitem[\protect\citeauthoryear{Kofman, Betschinger, Gelb, Nusser, Dekel}
{Kofman et al.}{1994}]{KBGND1994}
Kofman L., Bertschinger E., Gelb J.M. Nusser A., Dekel A,. 1994, ApJ, 420, 44
\bibitem[\protect\citeauthoryear{Kruse \& Schneider}{2000}]{KS2000} 
Kruse  G., Schneider, P., 2000, MNRAS, 318, 321
\bibitem[\protect\citeauthoryear{Ma \& Fry}{2000}]{MF2000} 
Ma C.-P., Fry J.N., 2000, ApJ, 543, 503 
\bibitem{M99} Mellier Y., 1999, ARA\&A, 37, 127
\bibitem[\protect\citeauthoryear{Miyazaki, Hamana, Shimasaku, Furusawa, Doi, Hamabe, Imi, Kimura, Komiyama, Nakata, Okada, Okamura, Ouchi, Sekiguchi, Yagi \& Yasuda}{Miyazaki et al.}{2002}]{MHSFDHIKKNOOOSYY2002} 
Miyazaki S., Hamana T., Shimasaku K., Furusawa H., Doi M., Hamabe M., Imi K., Kimura M., Komiyama Y., Nakata F., Okada N., Okamura S., Ouchi M., Sekiguchi M., Yagi M., Yasuda N., 2002, ApJL submitted
\bibitem[\protect\citeauthoryear{Moore, Quinn, Governato, Stadel \& Lake}
{Moore et al.}{1999}]{MQGSL1999}
Moore B., Quinn T., Governato F., Stadel J., Lake G., 1999, MNRAS, 310, 1147
\bibitem[\protect\citeauthoryear{Munshi \& Jain}{2000}]{MJ2000} 
Munshi D., Jain B., 2000, MNRAS, 318, 109
\bibitem[\protect\citeauthoryear{Nakamura \& Suto}{1997}]{NS1997} 
Nakamura T.T., Suto Y., 1997, Prog.Theor.Phys., 97, 49
\bibitem[\protect\citeauthoryear{Navarro, Frenk \& White}{Navarro et al.}{1996}]{NFW1996} Navarro J., Frenk C., White S. D. M., 1996, ApJ, 462, 564
\bibitem[\protect\citeauthoryear{Navarro, Frenk \& White}{Navarro et al.}{1997}]{NFW1997} Navarro J., Frenk C., White S. D. M., 1997, ApJ, 490, 493
\bibitem[\protect\citeauthoryear{Peacock \& Dodds}{1996}]{PD1996} 
Peacock J.A., Dodds S.J., 1996, MNRAS, 280, L19
\bibitem[\protect\citeauthoryear{Peebles}{1980}]{P1980} 
Peebles, P.J.E., 1980, {\it The Large-Scale Structure of the Universe} 
(Princeton Univ. Press)
\bibitem[\protect\citeauthoryear{Press \& Schechter}{1974}]{PS1974} 
Press W. H., Schechter P., 1974, ApJ, 187, 425
\bibitem{RKBS99}Reblinsky K., Kruse G., jain B., Schneider P., 1999, 
A\&A 351, 815
\bibitem[\protect\citeauthoryear{Scoccimarro, Sheth, Hui \& Jain}{Scoccimarro et al. 2001}{2001}]{SSHJ2001} 
Scoccimarro R., Sheth R. K., Hui L., Jain, B., 2001, ApJ, 546, 652
\bibitem[\protect\citeauthoryear{Seljak}{2000}]{S2000} 
Seljak U., 2000, MNRAS, 318, 203
\bibitem[\protect\citeauthoryear{Sheth \& Tormen}{1999}]{ST1999} 
Sheth R. K., Tormen G., 1999, MNRAS, 308, 119
\bibitem[\protect\citeauthoryear{Takada \& Jain}{2002}]{TJ2002} 
Takada M., Jain B., 2002, MNRAS, in press (astro-ph/0205055)
\bibitem[\protect\citeauthoryear{Taruya, Takada, Hamana, Kayo \& Futamase}
{Taruya et al.}{2002}]{TTHKF2002} 
Taruya A., Takada M., Hamana T., Kayo I., Futamase T., 2002, ApJ, 
571, 638
\bibitem[\protect\citeauthoryear{Taylor \& Watts}{2000}]{TW2000} 
Taylor A.N., Watts P.I.R., 2000, MNRAS, 314, 92 
\bibitem[\protect\citeauthoryear{Valageas}{2000}]{V2000} 
Valageas P., 2000, A\&A, 356, 771
\bibitem{WH93}Wu X.-P., Hammer F., 1993, MNRAS, 262, 187
\end{thebibliography}
\end{document}